\documentclass[preprint, 12pt]{emulateapj}

\usepackage{epsfig}
\usepackage{graphicx}
\usepackage{float}
\usepackage{xcolor}

\begin{document}


\title {X-ray Observations of Complex Temperature Structure in the Cool-core cluster Abell 85}
\author {David E. Schenck\altaffilmark{1} ,
         Abhirup Datta\altaffilmark{1,2},
	 Jack O. Burns\altaffilmark{1},
	 \and Sam Skillman\altaffilmark{3,4}
} 

\altaffiltext{1}{Center for Astrophysics and Space Astronomy, Department of Astrophysical and Planetary Science, University of Colorado, Boulder, CO 80309}
\altaffiltext{2}{NASA Postdoctoral Fellow}
\altaffiltext{3}{DOE Computational Science Graduate Fellow}
\altaffiltext{4}{Kavli Fellow, Kavli Institute for Particle Astrophysics and Cosmology, SLAC, CA 94025}


\begin{abstract}
X-ray observations were used to examine the complex temperature structure of Abell 
85, a cool-core galaxy cluster. Temperature features can provide evidence of merging 
events which shock heat the intracluster gas. Temperature maps were made from both 
\textit{Chandra} and \textit{XMM-Newton} obervations. The combination of a new, 
long-exposure \textit{XMM} observation and an improved temperature map binning 
technique produced the highest fidelity temperature maps of A85 to date. Hot 
regions were detected near the subclusters to the South and Southwest in both the 
\textit{Chandra} and \textit{XMM} temperature maps. The presence of these 
structures implies A85 is not relaxed. The hot regions may indicate the presence of
shocks. The Mach numbers were estimated to be $\sim$1.9 at the locations of the hot spots.
Observational effects will tend to systematically reduce temperature jumps,
so the measured Mach numbers are likely underestimated. Neither temperature map showed
evidence for a shock in the vicinity of the presumed radio relic near the Southwest
subcluster. However, the presence of a weak shock cannot be ruled out. There was tension
between the temperatures measured by the two instruments.
\end{abstract}

\keywords{galaxy clusters, x-ray, shocks}


\section{Introduction}

Galaxy clusters are formed gradually through the merging of structures within
the cosmic web. Major mergers can be highly energetic, releasing $\sim$
$10^{63}$-$10^{64}$ ergs of energy into the intracluster medium (ICM) (Kravtsov
\& Borgani 2012; Owers, Couch, \& Nulsen 2009; Kempner, Sarazin, \& Ricker
2002). These interactions thus have a dramatic effect on their surroundings,
leaving behind observable clues as to the presence and nature of merging
events. In order to obtain a complete description of the state of a cluster,
multi-wavelength data are necessary since the separate components of the
cluster are emitting through different mechanisms over a wide range of energy.

Galaxy clusters are classified as cool-core (CC) clusters or non-cool-core
(NCC) clusters based on the properties of the cluster core. Approximately half
of observed clusters are classified as CC clusters which tend to have denser,
cooler centers than their non-cool-core counterparts (Burns et al. 2008, Chen
et al. 2007). CC clusters are often believed to be more relaxed than
non-cool-core (NCC) clusters, making them more appealing for use in
cosmological studies since their thermal states are assumed to be not as
severely contaminated by interactions with their surroundings (Henning et al.
2009). Recent simulations challenge the belief that CC clusters are dynamically
relaxed systems by showing that the difference between CC and NCC clusters is
their early merger history. According to the simulations, NCC clusters may have
experienced a major merger, a merger in which the colliding objects are of
comparable mass, while CC clusters avoided such an event (Burns et al. 2008,
Henning et al. 2009). Major mergers may destroy early cool-cores before the
cluster is able to become cool and dense enough to withstand the interaction.
The simulations did not demonstrate any distinction between CC and NCC clusters
on the basis of their present day dynamical states. The distinction between CC
and NCC clusters is complicated by the fact that not all clusters fall cleanly
into one category or the other. Some clusters lie in between the two
categories; these clusters are referred to as weak cool-core (WCC) clusters by
Hudson et al. (2010). This work was partly motivated by the question of the
dynamical state of CC clusters.

Shocks created by the merging of clusters have important effects on the
evolution of structure in the cosmic web (Markevitch \& Vikhlinin 2007). Shocks
are responsible for heating the ICM by converting gravitational potential
energy into thermal energy and producing non-thermal populations of particles
which manifest as synchrotron and/or inverse-Compton emission (Skillman et al.
2008 and 2010, Ryu et al. 2003, Pfrommer et al. 2006). The shocks associated
with merging trace the formation of cosmological structure through the
radiation of heated gas. 

Observational evidence of shocks in clusters are
sometimes in the form of radio relics, structures of enhanced radio emission
which tend to be found in the outskirts of clusters and have arc-like
morphologies. They are believed to be the result of shocks induced by mergers,
compression of radio lobes, or the remnants of radio galaxies (Slee et al.
2001). Radio relics differ from radio halos which tend to be found near the
centers of clusters and have lower polarization and morphologies which match
the X-ray emission (Ferrari et al. 2008, Skillman et al. 2013). In principle,
shocks in the ICM should be associated with an X-ray excess and temperature
enhancement as shock heating should increase the emission
and heat the gas. \textit{XMM} observations of Abell 3667, a cluster undergoing
a major merger, revealed both an X-ray excess and temperature structure near
the Northwest radio relic (Finoguenov et al. 2010, Datta et al. 2014). However,
this is the largest relic detected to date; shocks in the intracluster medium
will not generally be detectable by their X-ray emission because they reside
greater than 1 Mpc from the cluster center where the X-ray surface brightness
is too low to reliably detect an enhancement (Hoeft \& Br{\"u}ggen 2007).

In this paper, we concentrate on Abell 85 (z=0.0555), which is classified as
a strong cool-core (CC) cluster based on the central cooling time (Hudson et
al. 2010). Abell 85 has been studied extensively in several wavelength regimes.
Spectral analysis has been conducted using X-ray observations by
\textit{Chandra} (Kempner, Sarazin, \& Ricker 2002; Lima Neto et al. 2003),
\textit{XMM-Newton} (Durret et al. 2005), \textit{Suzaku} (Tanaka et al. 2010),
\textit{ROSAT} (Durret et al. 1998), \textit{ASCA} (Markevitch et al. 1998),
and \textit{BeppoSAX} (Lima Neto, Pislar, \& Bagchi 2001). \textit{Chandra} and
\textit{XMM} have higher spatial and spectral resolution than the other
X-ray telescopes. Temperature maps created from these instruments are capable
of revealing the cool-core which has significantly lower temperature than the
surrounding gas. The X-ray observations also revealed two subclusters in the
vicinity of the main cluster, one located to the south and one to the
southwest.

Both the South and Southwest subclusters (see Figure 1) are associated with
radio sources. Abell 85 has been observed with the VLA at 333 MHz and 1.4 GHz.
The radio emission of the Southern subcluster may be due to a dying tailed
radio galaxy (Giovannini \& Feretti 2000). The Southwest subcluster is slightly
offset from a presumed radio relic (0038-096). The radio emission is located
$\sim$320 kpc from the cluster center in projection and has polarization as
high as 35\% (Slee et al. 2001). Tailed radio sources are located to the
northwest of the main cluster and the northeast of the Southern subcluster
(Giovannini \& Feretti 2000).

The galaxy population of A85 has been observed at optical wavelengths to study
their distribution and dynamics. Substructures were detected in the spatial and
redshift distributions. It was found that there are merging substructures to
the southeast of the main cluster which are not in the field of view of our
\textit{Chandra} observation (Bravo-Alfaro et al. 2009). The morphology and
star formation rate of such infalling galaxies change as a result of their
interaction with the main cluster (McIntosh, Rix, \& Caldwell 2008). The orbits
of the galaxies was shown to be consistent with being isotropic (Hwang \& Lee
2008).

The data used for this paper includes a 40 ks \textit{Chandra} observation from
2000 which has been used to create temperature maps (Kempner, Sarazin, \&
Ricker 2002; Lima Neto et al. 1998) as well as a new 100 ks \textit{XMM}
observation. \textit{XMM} had been used to observe A85 before, but the
exposure time was approximately 12 ks. The new observation provides
a significant improvement over existing X-ray observations because the longer
exposure allows for higher resolution temperature maps. The temperature maps
created using both \textit{Chandra} and \textit{XMM} were improved in this work
by applying a binning technique adapted from Randall et al. (2008). In this
work, we used the X-ray observations to find evidence of mergers through
shock-heating and to study the relationship between the X-ray temperature and
the extended radio emission in A85.

The paper is organized as follows: a summary of the archival data and their
reduction are given in Section 2. A detailed description of how the temperature
maps were made is presented in Section 3. This includes descriptions of the two
binning methods as well as a discussion on fitting X-ray spectra. The
temperature maps are presented in this section. Section 4 includes
a description of estimating Mach numbers under the assumption that hot spots in the
temperature maps result from shocks. Conclusions are presented in Section 5.

\section{Data and Reduction}

\subsection{X-ray Data Reduction}

We used both \textit{Chandra} and \textit{XMM-Newton} observations in our
analysis. The \textit{Chandra} observation (ID 0904) was obtained from the
Chandra Data Archive. It was taken in August 2000 with an exposure time of
$\sim$39 ks in FAINT mode. The \textit{Chandra} flux map is presented in Figure
1. The cool-core of A85 is offset from the center of the ACIS-I detector to
avoid having the core fall on the borders between the detectors which have
reduced sensitivity.

A new \textit{XMM} observation (ID 0723802201) was performed June 16, 2013, in
Full Frame mode with a MEDIUM optical filter for an exposure time of 100 ks.
Prior to June 2013, the best \textit{XMM} observation of A85 available was
0065140101 which had only 12 ks of exposure time. We also processed this
dataset, but we do not present it here; the more recent observation offered
much better resolution in the temperature maps because of the increased
exposure.

\begin{figure*}[t!]
  \centering
  \includegraphics[height=12cm]{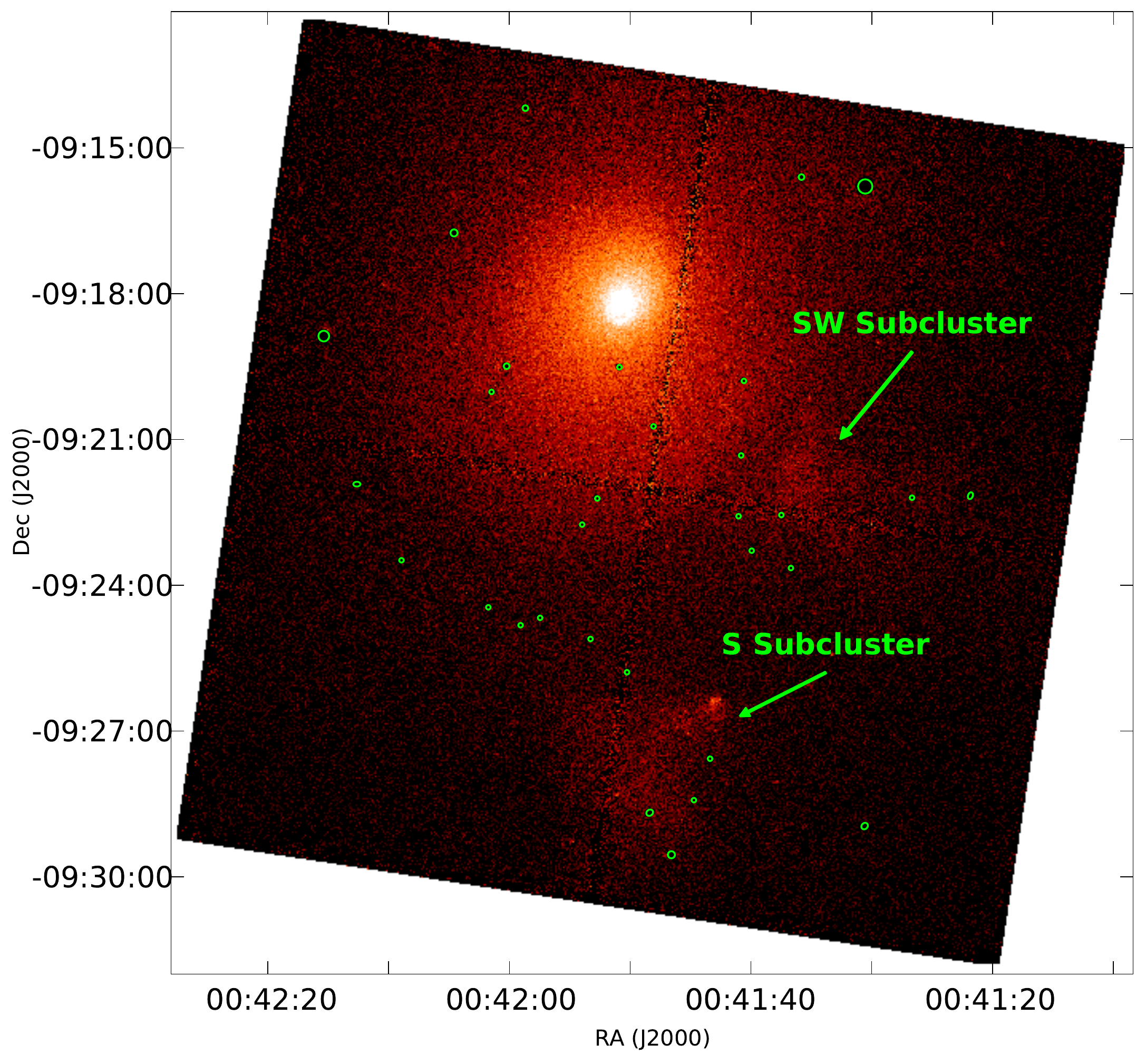}
  \caption{\textit{Chandra} flux map. The location of the Southern (S) and
Southwest (SW) subclusters are indicated as well as the locations of point
sources which were removed for spectral analysis.}
\end{figure*}

\begin{figure*}[t!]
  \centering \leavevmode
  \includegraphics[height=12cm]{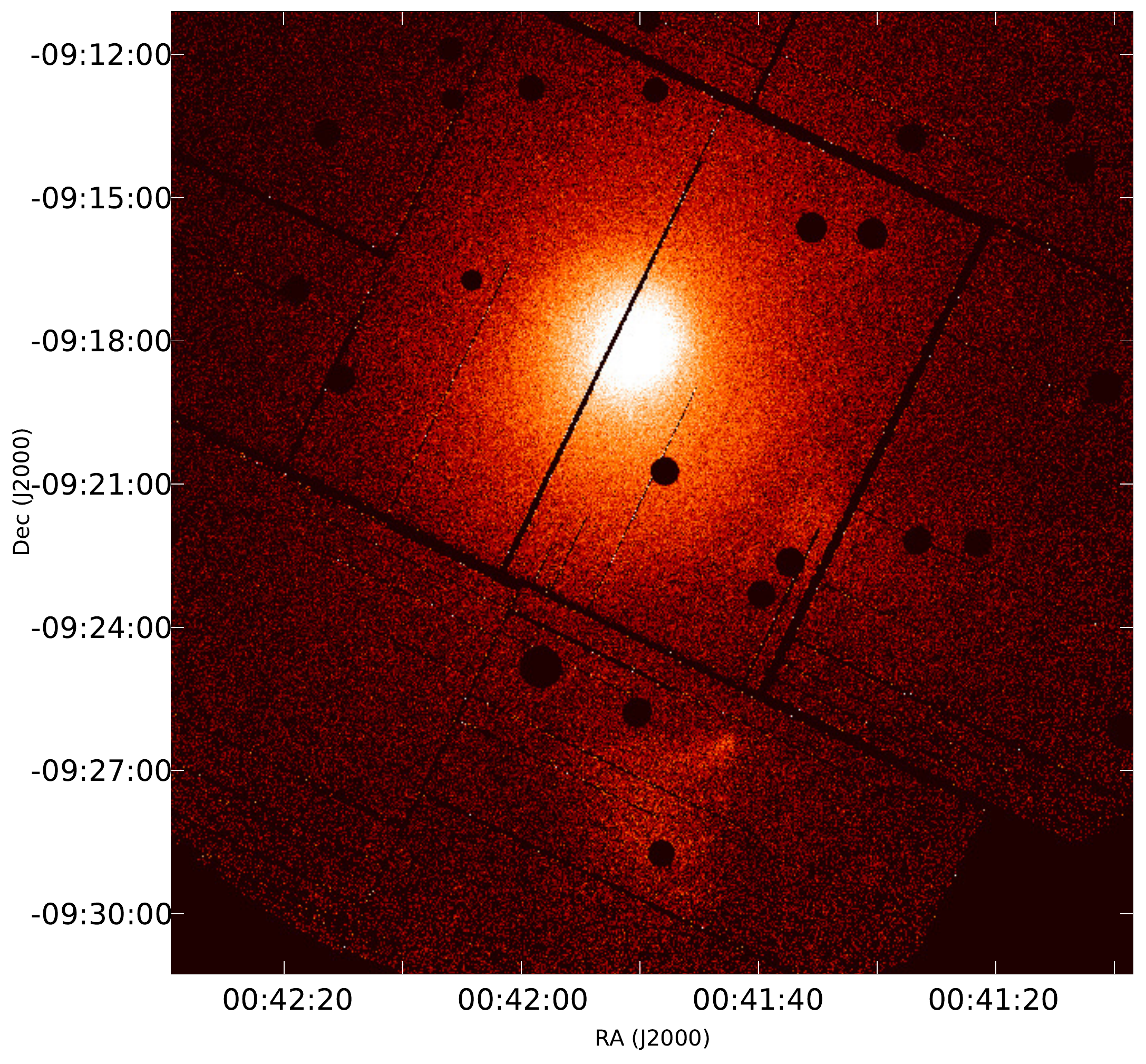}
  \caption{\textit{XMM} flux map. The point sources which were removed for
spectral analysis are clearly visible as holes in the data.}
\end{figure*}

The \textit{Chandra} data were calibrated using \textit{CIAO} 4.3 and
\textit{CALDB} 4.4.6, the most up-to-date versions at the time of analysis. Bad
pixels and cosmic rays were removed using
\textit{acis}\_\textit{remove}\_\textit{hotpix} and CTI corrections were made
using \textit{acis}\_\textit{process}\_\textit{events}. Intervals of background
flaring were excluded using light curves in the full band and 9-12 keV band.
The light curves were binned at 259 seconds per bin, the binning used for the
blank-sky backgrounds. Count rates greater than 3-$\sigma$ from the mean were
removed using \textit{deflare}. We visually inspected the light curves to
ensure flares were effectively removed. We used the blank-sky backgrounds in
CALDB 4.4.6. The backgrounds were reprojected and processed to match the
observations.

\begin{deluxetable}{cccc} \tablecolumns{4} \tabletypesize{\footnotesize}
  \tablecaption{X-ray data summary} \tablewidth{0pt} \tablehead{
    \colhead{Telescope} & \colhead{ID} & \colhead{Exposure [ks]}
    & \colhead{Date} } \startdata XMM	&	0723802201	&	100	&	6/2013	\\
    Chandra	&	0904	&	39	&	8/2000	\\ \enddata \tablecomments{Summary
    of observations used for X-ray analysis.} \end{deluxetable}

Point sources were removed to prevent contamination of the spectra. Point sources
were identified using \textit{wavdetect} on counts in the 0.2-12 keV band,
though manual inspection was necessary as \textit{wavdetect} made many false
identifications and failed to identify all real sources. We masked the regions
containing point sources from both the cluster observation and the background
to prevent over-subtracting the background. The locations of removed point
sources are indicated in Figure 1 and are apparent in Figure 2.

The steps above were also performed on the new 100 ks \textit{XMM} observation
using \textit{SAS} version 11.0.0. Event files were created using
\textit{emchain} and filtered for bad pixels, bad columns, and cosmic rays
using \textit{evselect}. Point sources are detected using \textit{eboxdetect}
and, as with \textit{Chandra}, the results were manually inspected and adjusted
to account for the imperfect identifications made by the software. The point sources
are detected in the 0.2-12 keV band. Time intervals with flares were removed by
plotting the light curves and identifying the intervals to ignore. We used
backgrounds obtained from the \textit{XMM-Newton EPIC} background working group at the
University of Leicester (Carter et al. 2007).

\begin{figure*}[t!]
  \centering
  \includegraphics[height=10.16cm]{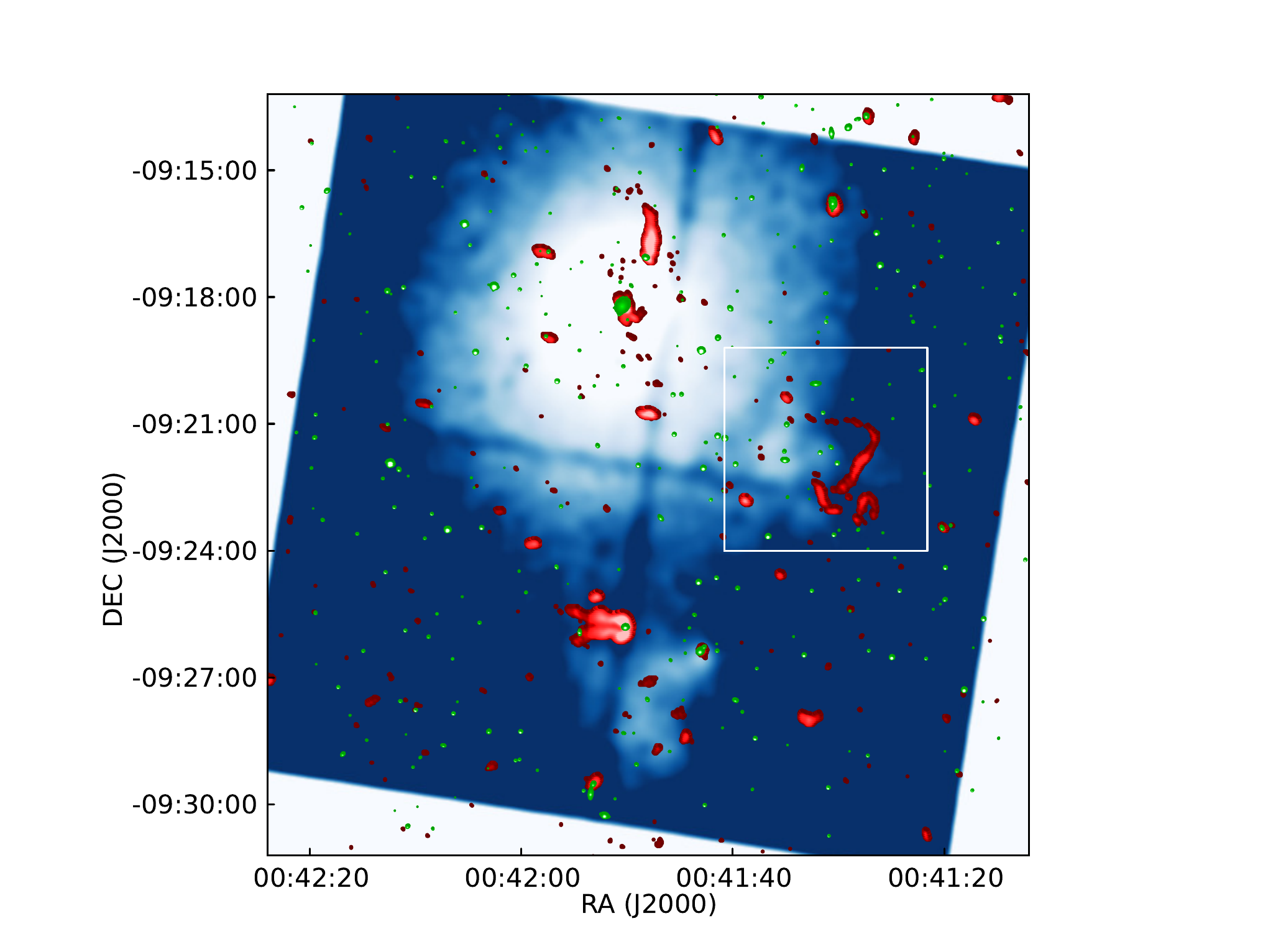}
  \includegraphics[height=10.16cm]{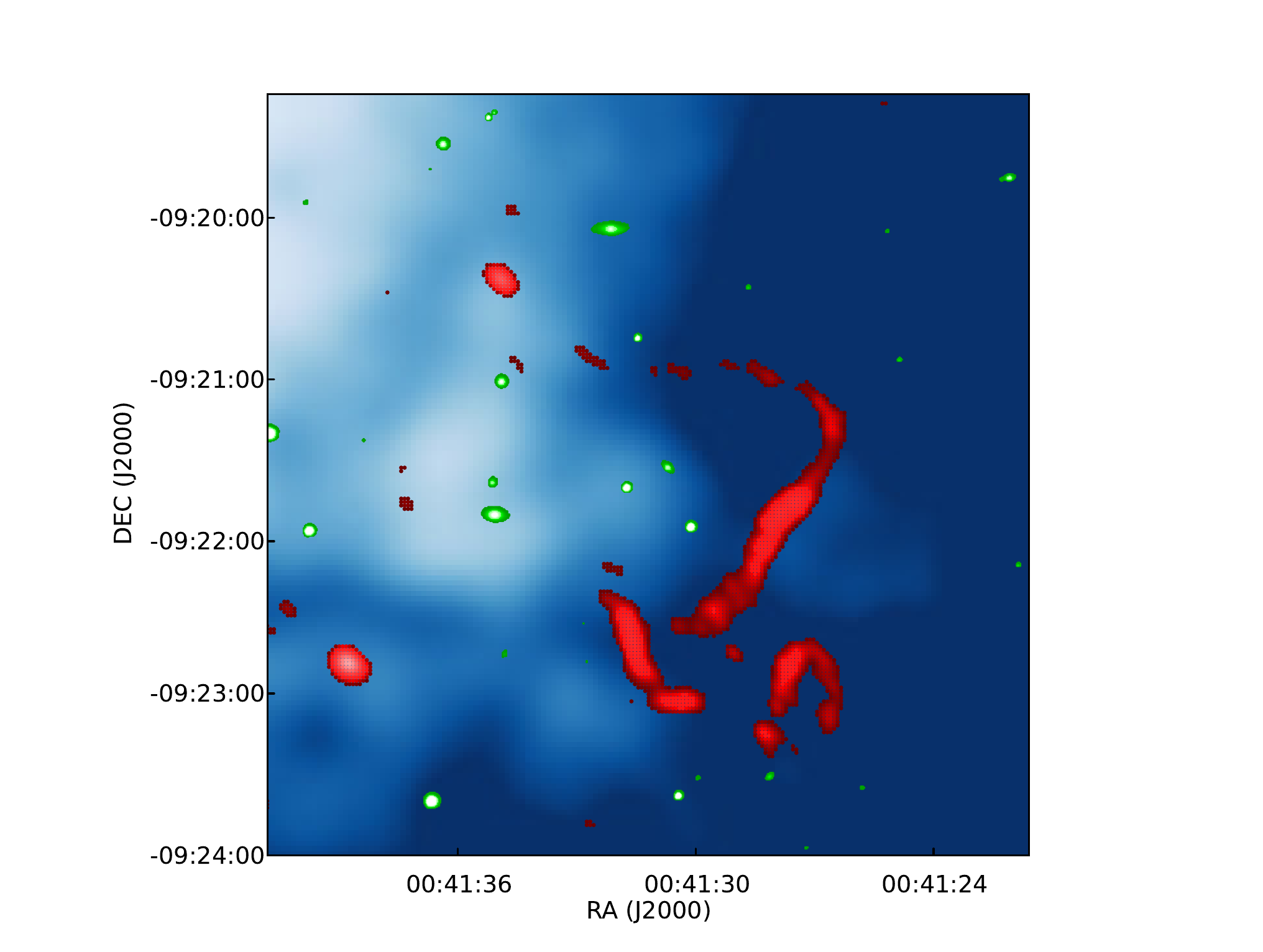}
  \caption{Top: X-ray flux measured by \textit{Chandra} with 1.4 GHz radio
(VLA) overlayed in red and r-band optical (SDSS) overlayed in green. Bottom:
Same as the top panel, zoomed in to the region enclosed by the white square.
The presumed radio relic is shown in the red. It has a thin, filamentary
structure and it is located to the west of the Southwest subcluster.}
\end{figure*}

\subsection{Radio Data Reduction}

To obtain a more complete picture of A85, we also examined archival optical and
radio data. Figure 3 shows an overlay of X-ray, optical, and radio. 

We reanalyzed the radio data presented in Slee et al. (2001) using the
Astronomical Image Processing System (AIPS). The data were taken by the VLA in
the B configuration on September 1, 2001 at a frequency of 1.4 GHz. Three
dimensional imaging was done using 31 facets and weighting parameter of robust
value equal to one. The final image, shown in red in Figure 3, was made after
several rounds of imaging and self-calibration of the phase. The restoring beam
size of the image is 7.7$\times$5.4 arcseconds. The off-source RMS of the final
image is 15.3 $\mu$Jy/beam and the on-source RMS near the radio emission to the
Southwest is 44.8 $\mu$Jy/beam. It should be noted that our current image has
improved noise properties compared to the previously published image (Slee et
al. 2001). 

The optical image, shown in green, is an r-band image obtained from the Sloan
Digital Sky Survey (SDSS). The SDSS image does not show any galaxies coincident
with the radio emission.

\section{X-ray Spectral Analysis}

We created temperature maps by fitting a thermal plasma model to X-ray spectra.
For \textit{Chandra}, the source and background spectra were extracted using
\textit{dmextract} and the weighted response was extracted using
\textit{specextract}. The analogous \textit{SAS} commands for extracting
\textit{XMM} spectra and responses are \textit{evselect}, \textit{rmfgen}, and
\textit{arfgen}. We rescaled the background spectra using the ratio of the
high-energy counts (9.5-12 keV) in the source and the background. This corrects
for the fact that the background might have been stronger or weaker at the time
of the observation compared to the blank-sky background file. We expect the
counts to be predominantly from the background at these high energies.

Two different methods were used to construct the temperature maps. The
procedure for each method, as well as the pros and cons of each, are described
here.

The bottom-left panels of Figures 4 and 6 were produced using a weighted
Voronoi tesselation (WVT) routine (Diehl \& Statler 2006) which defines regions
which have approximately the same signal-to-noise ratio. The signal is equal to
the background subtracted counts and the noise assumes Poisson contributions
from both the source and background. The background was rescaled by the ratio
of the source and background exposure times for the sake of WVT binning. The
result of the WVT binning routine is a map divided into distinct, roughly
polygon-shaped regions that are smaller where the signal is higher and larger
where the signal is lower. The signal-to-noise threshold is set by the user; we
converged on a threshold of 50 which tends to result in approximately 10$\%$
errors on temperature. Using this threshold, our \textit{Chandra} map is
comprised of 112 regions and our \textit{XMM} map is comprised of 561 regions.
There are more regions in the \textit{XMM} temperature map because of the higher
integration time, larger field of view, and higher effective area.

The top-left panels of Figures 4 and 6 were produced using a method adapted
from Randall et al. (2008) and Randall et al. (2010) which we call Adaptive
Circular Binning (ACB). In these two papers, spectra were extracted from
circular regions which were just large enough to reach some threshold of
counts. The fitted temperature of each region was assigned to the pixel at
the center of the circle. This could be done for every pixel, but it is usually
done for every few pixels to save on time. The circles are allowed to overlap,
so some pixels will share counts with other pixels and the fitted temperatures
will not be independent from one another. For our work, we replaced the counts
threshold with a signal-to-noise threshold similar to the WVT binning. This
means the circle at some location should have about the same area as a WVT
region at the same location. The cost of extracting spectra from the overlapping
circles is that the number of spectra rises dramatically. Using a signal-to-noise
threshold of 50 and after binning the map into 6$\times$6 arcsecond pixels, the
number of spectra for \textit{Chandra} was 30,287. For \textit{XMM}, there are
31,417 ACB circles after binning to 6$\times$6 arcsecond pixels.

The WVT and ACB methods each have their advantages and shortcomings. The WVT
method creates regions which are independent from one another, but the ACB
method has overlapping extraction regions, making it a little more difficult to
interpret the temperature structures. The weakness of the WVT method is that
the regions might not line up optimally with the true underlying temperature
structure. The ACB method does not have this problem, so it is considered more
useful for revealing temperature structure.

The spectra were fit using an \textit{APEC} thermal plasma model in
\textit{XSPEC}. We included photoelectric absorption, but the Galactic hydrogen
column density was not fit. Instead, it was frozen at a value of 2.8$\times$
$10^{20}$ cm$^{-2}$ (Kempner et al. 2002). The temperature and metallicity were
fit for each spectrum, though in some cases the metallicity was not well
constrained and the fitted value was unreasonably high or low. For the WVT
temperature map, we corrected these cases by fixing the metallicity at 0.3
times solar, a reasonable value for the cluster as a whole. For the ACB map,
the fits were repeated with the metallicity fixed at a value determined from
metallicity fits in nearby regions. The C-statistic was used for all fits (Cash 1979).
Uncertainty was estimated using the Monte-Carlo technique in \textit{XSPEC}. To
save on computation time, the uncertainty was only calculated for the WVT
temperature map.

In the two sections below, the temperature maps are described for
\textit{Chandra} and \textit{XMM} separately. They are then compared and the
attempts to erase the discrepancies between the two are discussed.

\subsection{\textit{Chandra}}

The ACB and WVT \textit{Chandra} temperature maps are displayed in Figure 4.
The cool-core, located at RA 00:41:48 and Dec -09:18:00, reaches a low
temperature of 2.7 keV. To the Southeast of the main cluster, there is an
extended region of high temperature ranging from 8 to 14 keV. Another hot spot
is located between the main cluster and the Southwest subcluster which reaches
10.8 keV.

\begin{figure*}[t!]
  \centering
  \includegraphics[height=7.11cm]{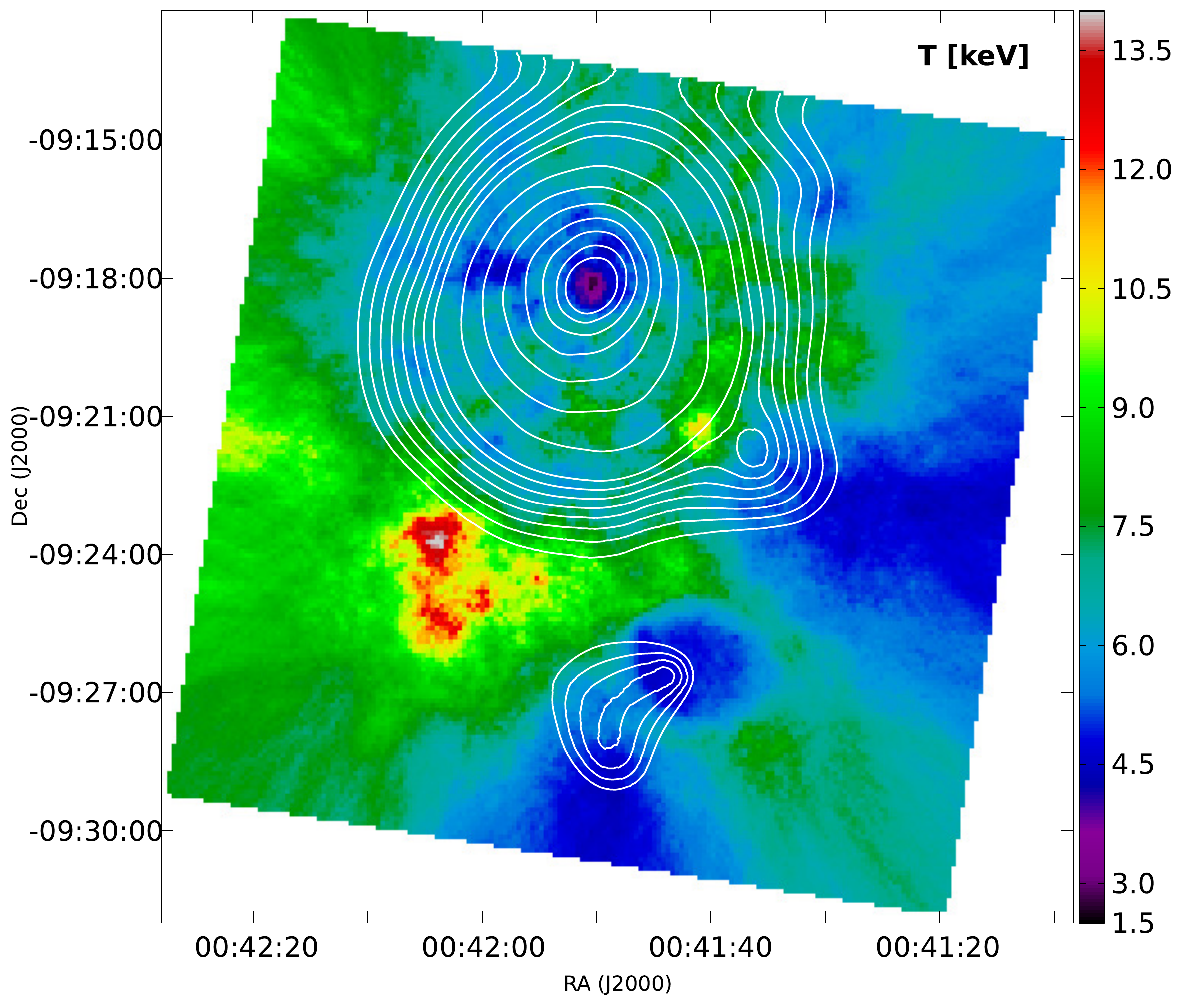}
  \includegraphics[height=7.11cm]{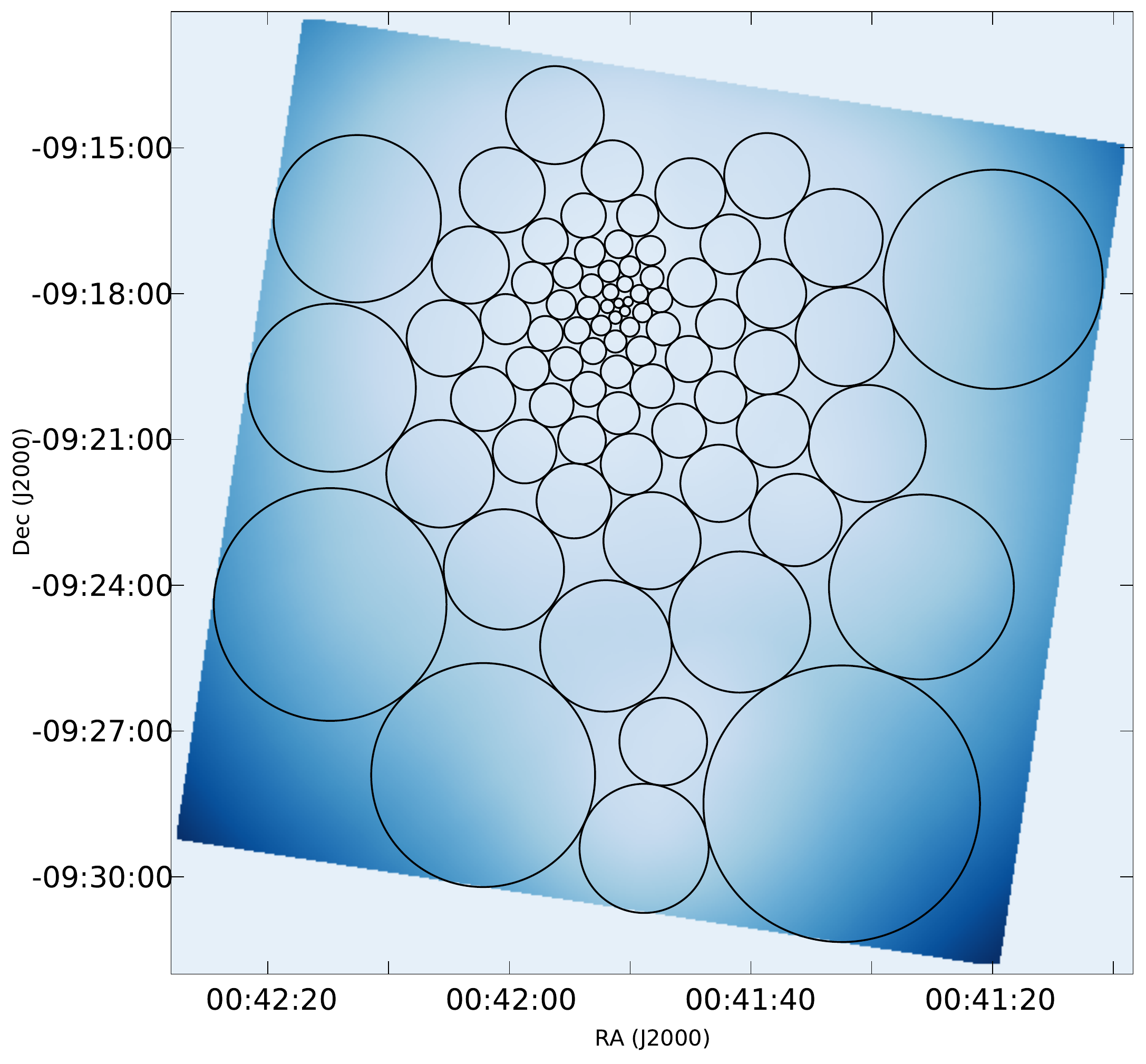}
  \includegraphics[height=7.11cm]{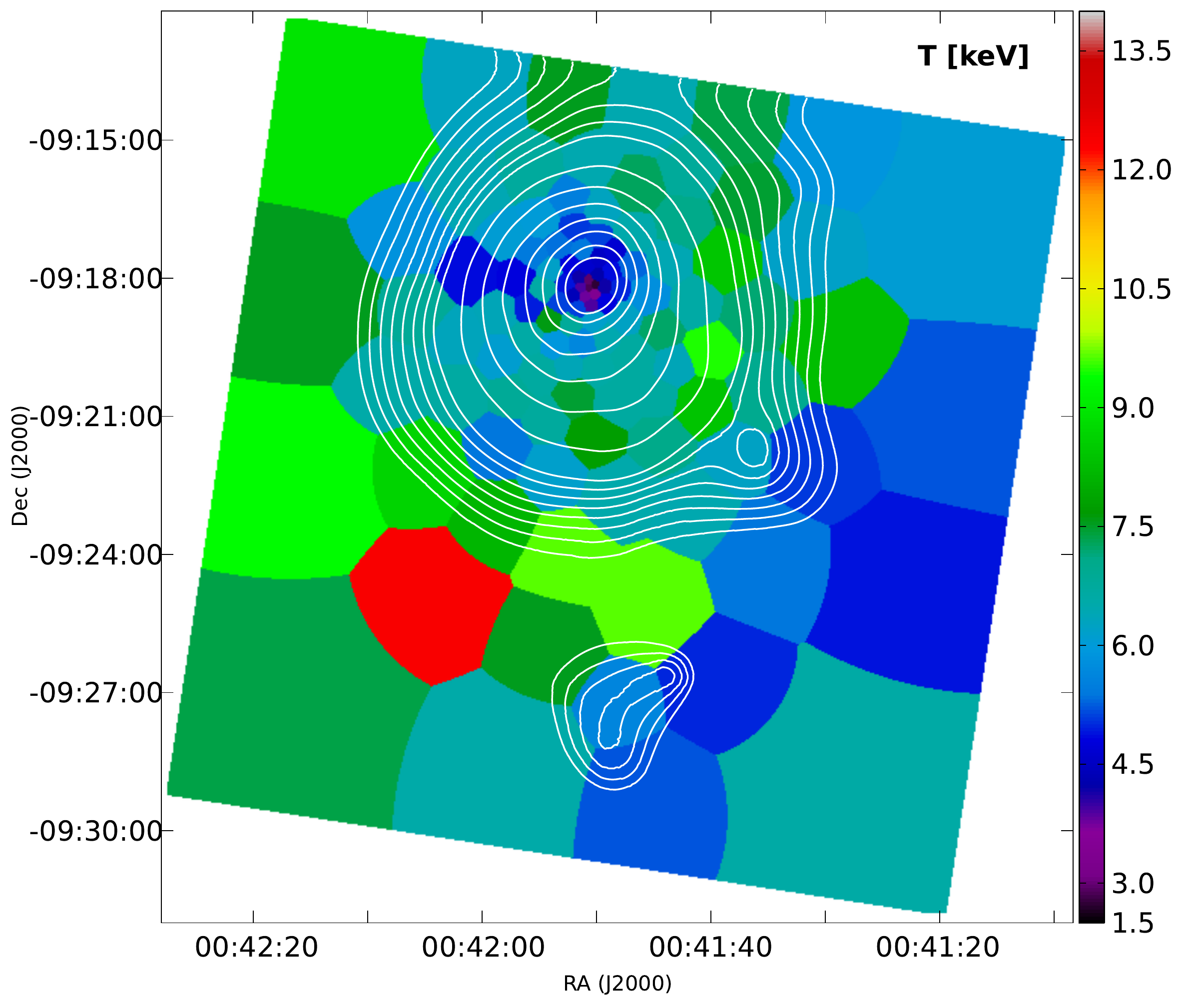}
  \includegraphics[height=7.11cm]{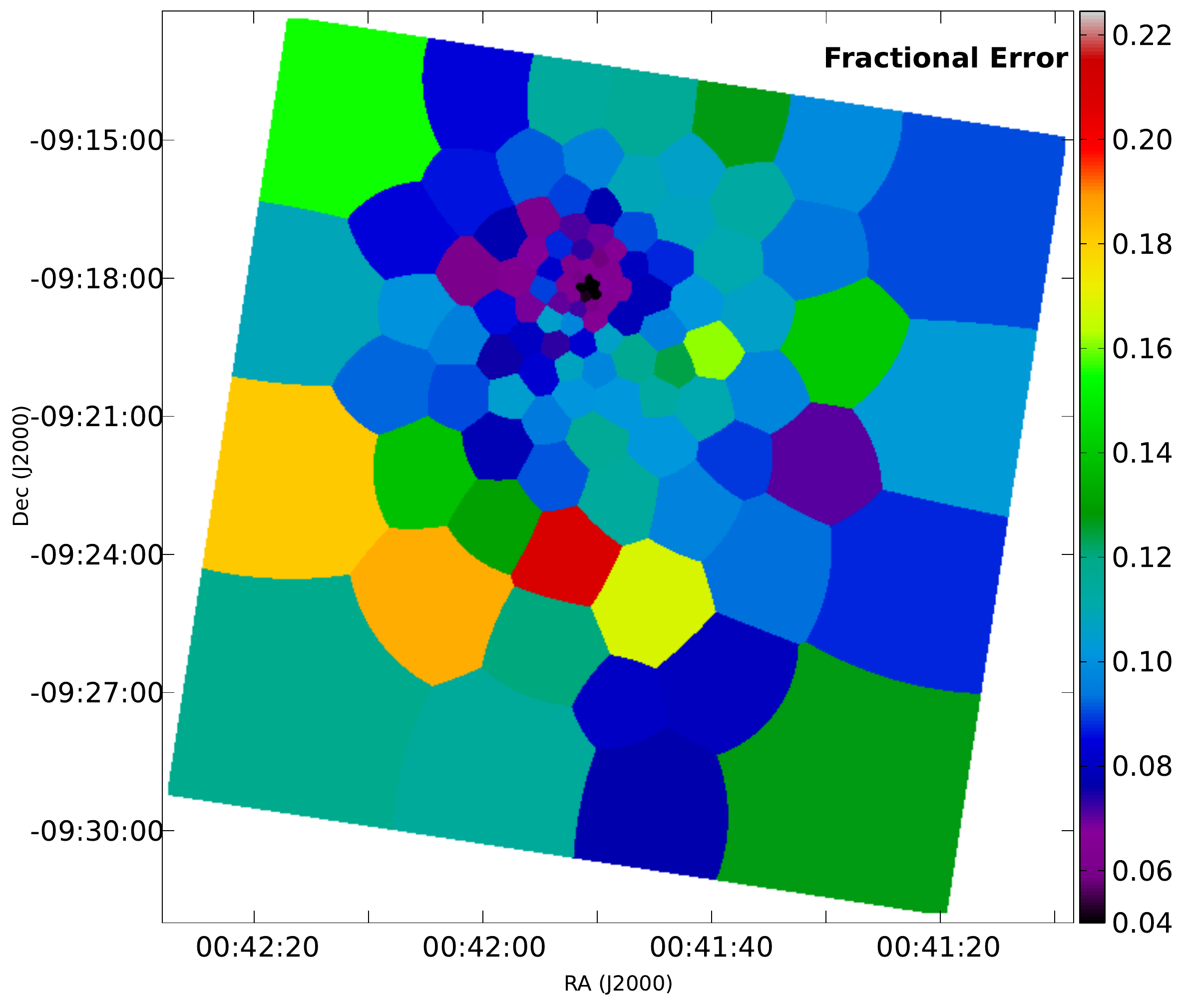}
  \caption{Top Left: \textit{Chandra} ACB temperature map. The contours 
are X-ray surface brightness. Top Right: Map of scales demonstrating the 
size of ACB extraction regions. The circles show a subset of the extraction 
regions. Bottom Left: \textit{Chandra} WVT temperature map. The contours 
are X-ray surface brightness. Bottom Right: \textit{Chandra} WVT temperature 
fractional error map.}
\end{figure*}

The bottom right panel of Figure 4 shows fractional errors calculated as

\begin{equation} \sigma_{frac} = \frac{\sigma_{+} + \sigma_{-}}{2T}
\end{equation} where $\sigma_{+}$ and $\sigma_{-}$ are the plus and minus
one-sigma errors calculated from the Monte Carlo error estimation in
\textit{XSPEC}. Using the Monte Carlo error estimation is time consuming, so it
was only run for the WVT temperature maps. The uncertainties from the WVT
\textit{Chandra} map are shown in the left panel of Figure 5. The typical
fractional error is $\sim$10$\%$.

\begin{figure*}[t!]
  \centering
  \includegraphics[height=6.35cm]{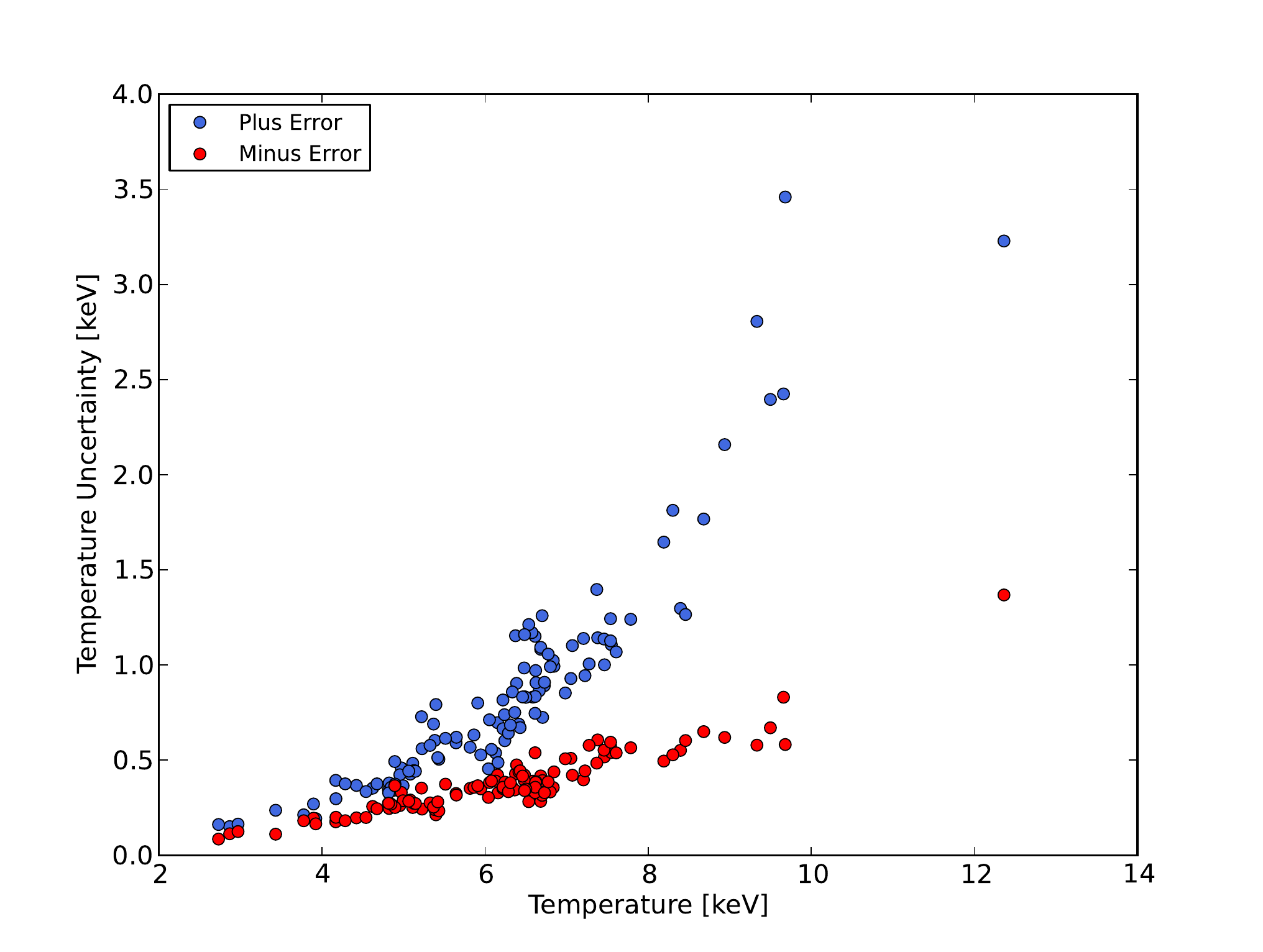}
  \includegraphics[height=6.35cm]{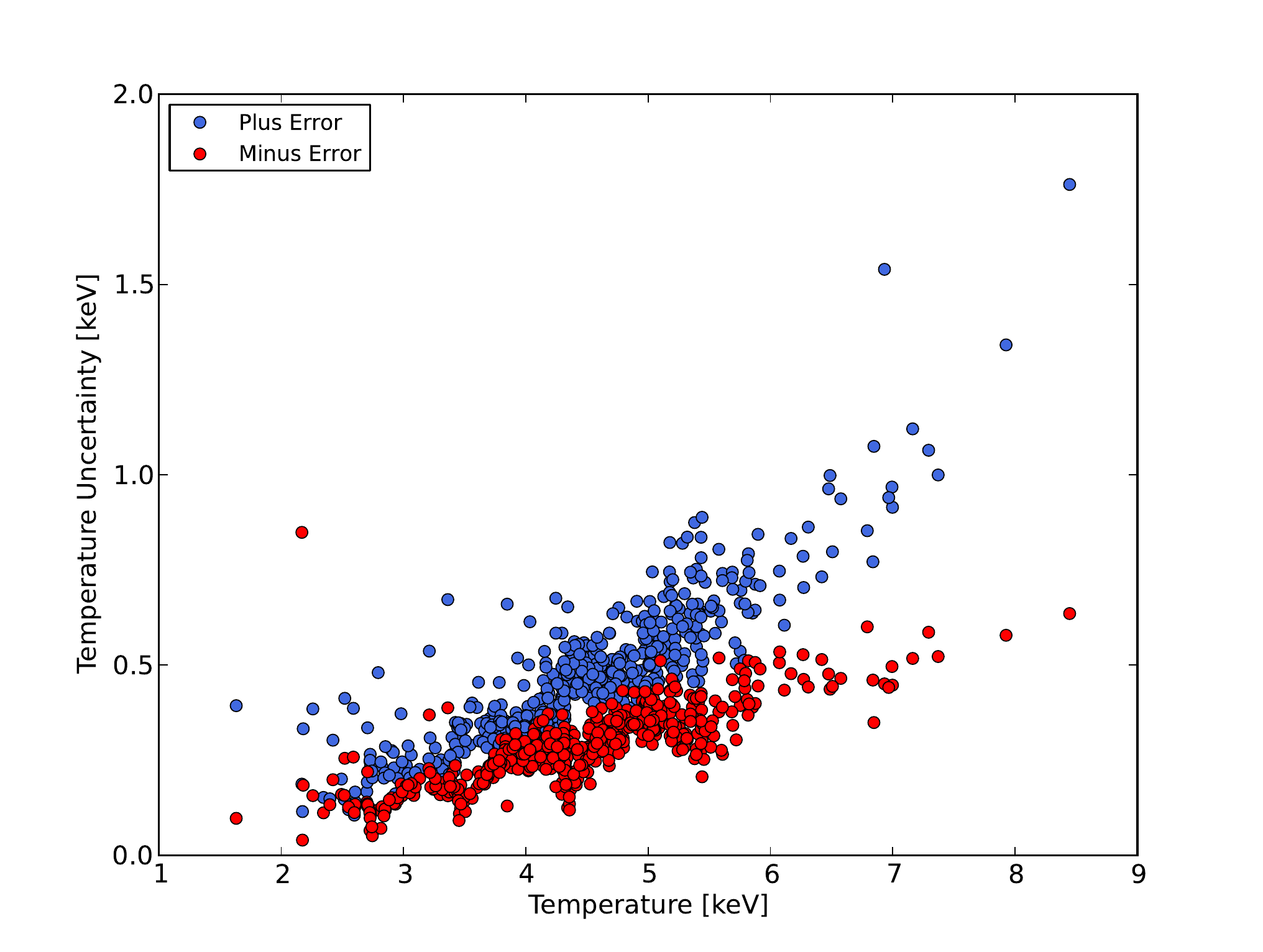}
  \caption{Left: Temperature uncertainties for \textit{Chandra} WVT map. 
Right: Temperature uncertainties for \textit{XMM} WVT map. In both cases, 
the plus errors tend to be higher than the minus errors.}
\end{figure*}

Near the edges of the \textit{Chandra} temperature map, there appear to be
streaked features which point approximately radially. The ACB circles grow as
the distance from the cluster center increases, as evidenced by the top-right
panel of Figure 4. Circles become especially large very close to the detector
edge because much of the circle encloses regions not on the detector. The
diameter of the largest circle is 80$\%$ of the width of the detector. These
very large circles expand inward toward the center of the map, resulting in
a smearing near the detector edge.

\subsection{\textit{XMM-Newton}}

The \textit{XMM} temperature maps are presented in Figure 6. Both \textit{MOS}
detectors were used to create temperature maps. The \textit{XMM} observation
had about 2.5 times the exposure as \textit{Chandra}. This combined
with the higher effective area of \textit{XMM} resulted in smaller extraction
regions than in the \textit{Chandra} map. Even with the improved resolution on
the WVT temperature map, the ACB map performs much better at revealing temperature
features.

\begin{figure*}[t!]
  \centering
  \includegraphics[height=7.11cm]{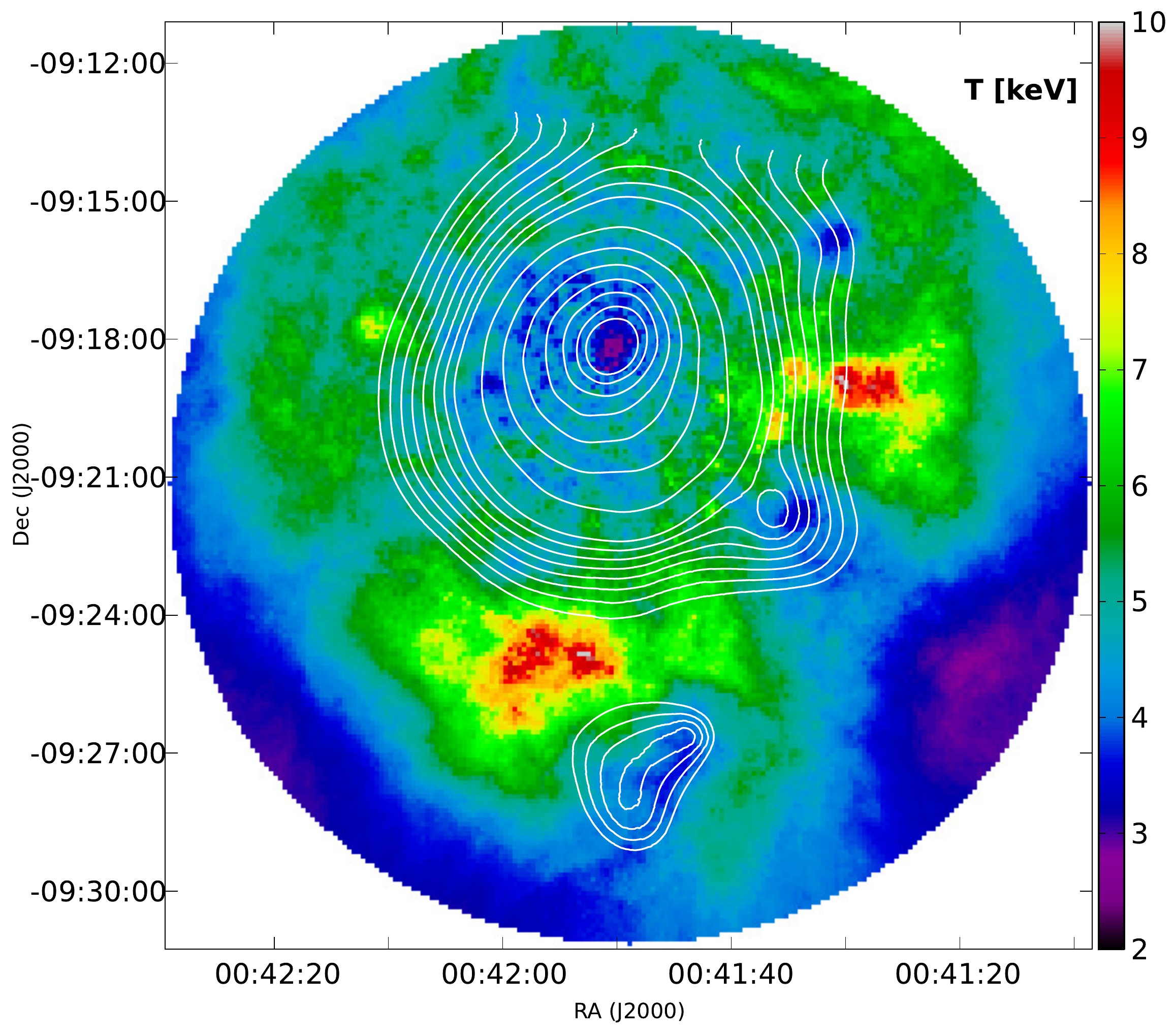}
  \includegraphics[height=7.11cm]{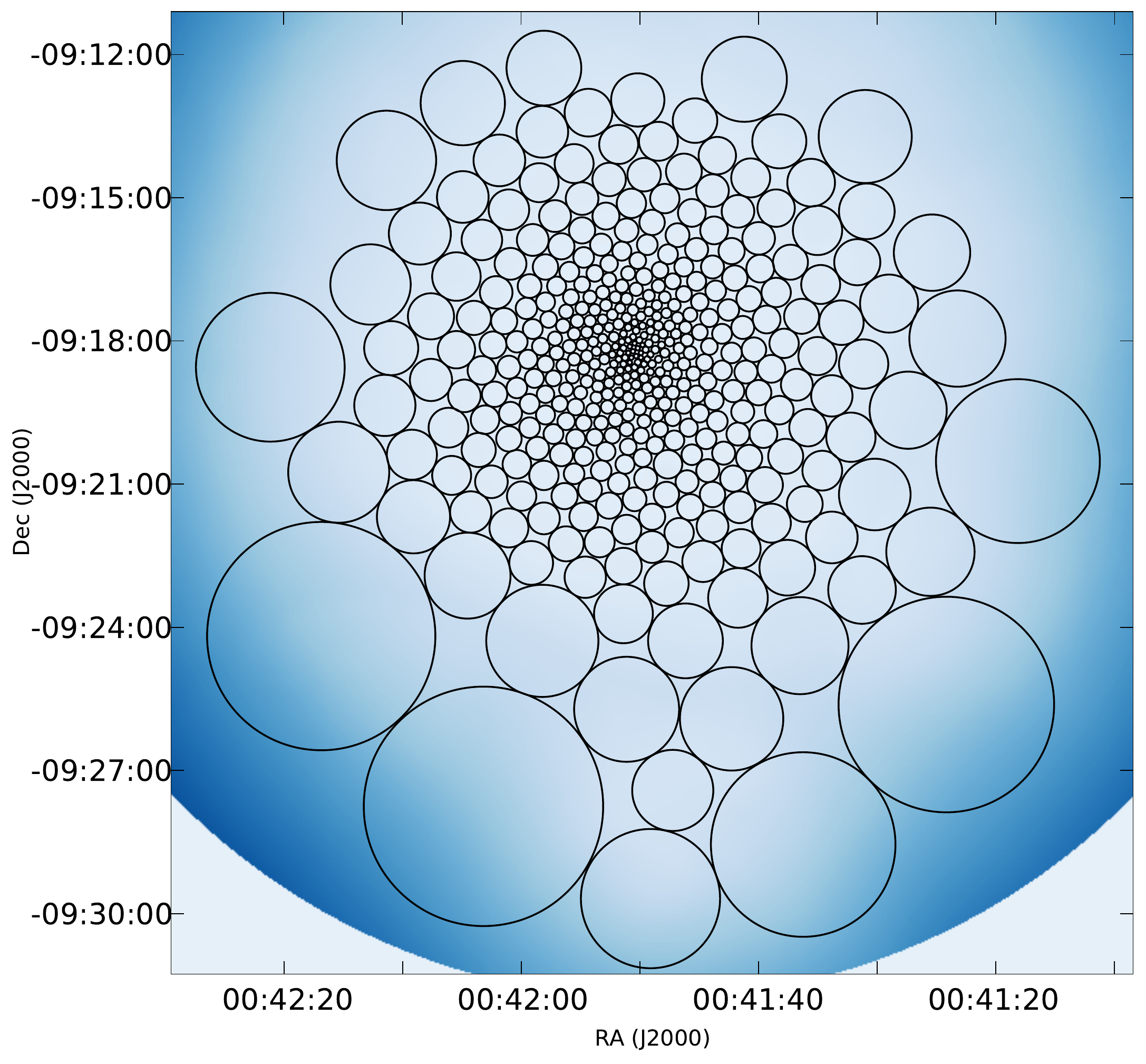}
  \includegraphics[height=7.11cm]{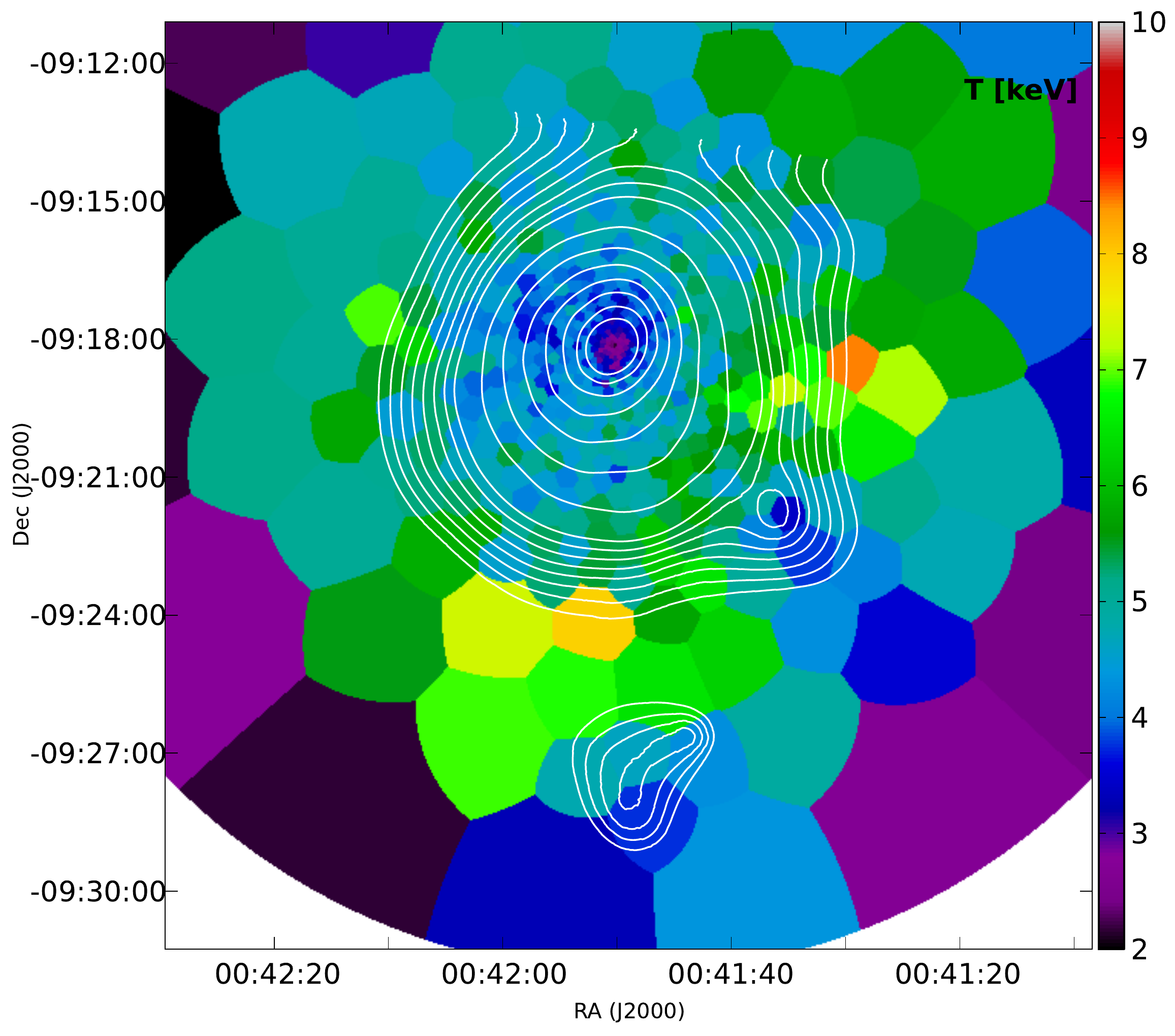}
  \includegraphics[height=7.11cm]{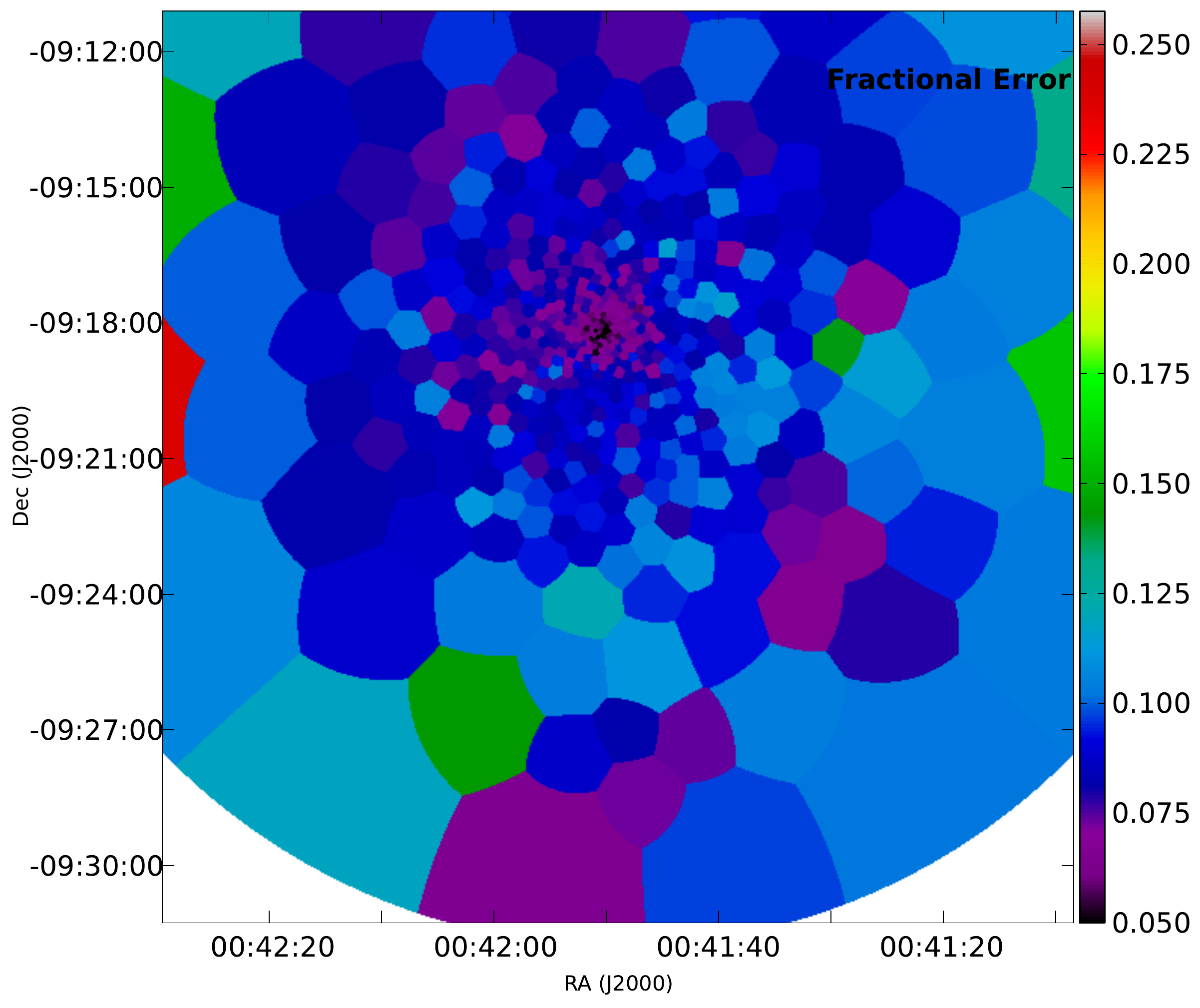}
  \caption{Top Left: \textit{XMM} ACB temperature map. The contours are 
X-ray surface brightness. Top Right: Bottom Left: Map of scales demonstrating 
the size of ACB extraction regions. The circles show a subset of the 
extraction regions. \textit{XMM} WVT temperature map. The contours are X-ray 
surface brightness. Bottom Right: \textit{XMM} WVT temperature fractional 
error map.}
\end{figure*}

The average temperature throughout the map is approximately 5 keV. The lowest
temperature in the map is reached within the cool-core which drops to 2.5 keV.
Just as with the \textit{Chandra} map, there is a region of hot gas located to
the northeast of the Southern subcluster. The peak temperature in this region
is 10.2 keV. The temperature also peaks at 10.2 keV to the north of the
Southwest subcluster.

The right panel of Figure 5 shows the plus and minus uncertainties for the
\textit{XMM} temperatures. The uncertainties are similar to those for
\textit{Chandra} for temperatures below $\sim$6 keV, but are typically lower
for higher temperatures.

The streaked temperature features discussed in the previous section do not
appear in the \textit{XMM} temperature map. To save time on fitting spectra,
only a subsection of the full \textit{XMM} field-of-view was utilized for
creating temperature maps. By excluding pixels close to the edge of the
detector, the circles were not allowed to grow to such large sizes. This
prevented the edges of the \textit{XMM} temperature map from suffering from
the same problem seen in the \textit{Chandra} map.

\subsection{Temperature Structure in an \textit{Enzo} Simulation}

Both the \textit{Chandra} and the \textit{XMM} temperature maps include an
extended region of hot gas offset to the Northeast of the Southern subcluster.
The Southern subcluster is believed to be falling in toward the main cluster
with a non-zero impact parameter. It is estimated that the subcluster will pass
within approximately 580 kpc to the West of the cluster center (Kempner,
Sarazin, \& Ricker 2002). The subcluster appears to be traveling to the
Northwest. The bright galaxy resides at the Northwest edge of the subcluster,
possibly because the gas is impeded by ram pressure exerted by the main cluster
ICM.

There are two features of the temperature structure near the Southern
subcluster that we seek to explain. One is that the temperature is not
symmetric about the supposed infall direction of the subcluster. If the hot gas
results from shock heating by the subcluster, there should be heated gas on
both sides of the subcluster in projection. However, the hot gas to the
Northeast is not matched by hot gas to the West. The second feature is the
extent of the heated region. The heated gas extends over an area far from the
Southern subcluster.

Interpreting these features is made difficult by the fact that observations of
galaxy clusters are effectively just snapshots in time and the
three-dimensional structure is compressed into a two-dimensional
representation. These limitations make formulating the exact state of a cluster
very difficult. Simulations do not suffer from these limitations. Both the time
evolution and full spatial information are available, making it far simpler to
characterize the dynamical state of a cluster. While cosmological simulations
might not be perfect representations of actual clusters, they do provide a more
detailed view of how clusters and subclusters interact and how such
interactions affect the thermal state of a cluster. Thus, cluster simulations
can be a valuable tool for interpreting observations of galaxy clusters.

To help illustrate how the temperature structure in A85 may have arisen, we
present a simulated cluster which qualitatively matches the interaction between
A85 and the Southern subcluster. The cluster was chosen from a cosmological
simulation including 80 clusters which were presented by Skory et al. (2013).
Each simulated cluster underwent multiple mergers which encompassed a wide
range of impact parameter, cluster mass ratio, velocity, and orientation
relative to the line-of-sight.

A snapshot of the simulated cluster temperature is presented in Figure 7. The
temperature was weighted by the X-ray emissivity. A subcluster is falling
toward the main cluster from the South, shock heating the gas ahead of it. The
shock heated gas is not distributed symmetrically about the infall direction of
the subcluster which is indicated by the white arrow. To the West, the shock
creates a narrow layer of hot gas, but to the East, the hot gas resides in
a large region which extends far from the subcluster. The extended region of
hot gas resembles that in A85 which is also disconnected from the subcluster
and covers a large area.

\begin{figure*}[t!]
  \centering
  \includegraphics[height=10.16cm]{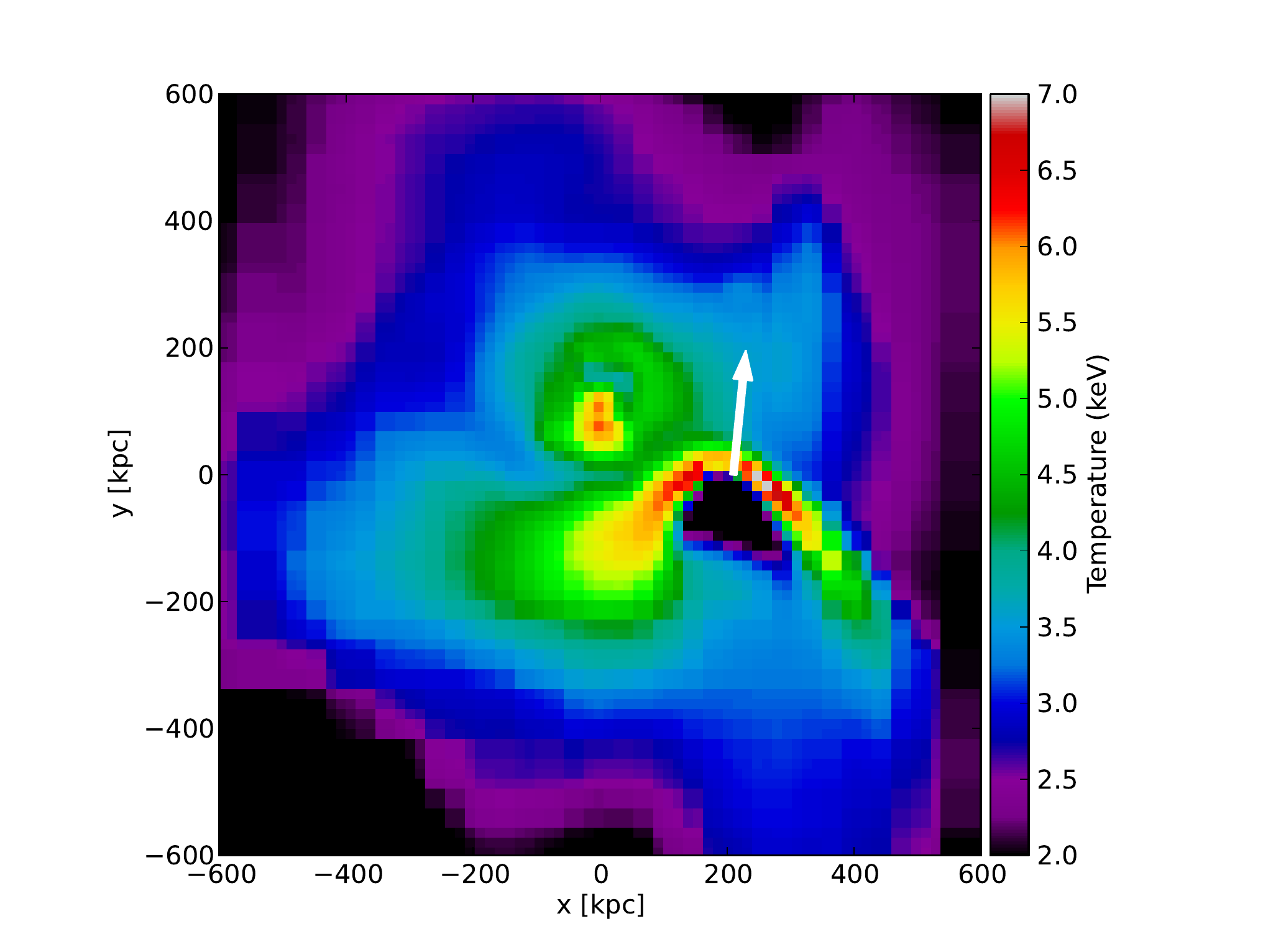}
  \caption{Temperature map of a single galaxy cluster produced in \textit{Enzo}. 
The subcluster is moving in a direction indicated by the white arrow. A shock 
precedes the subcluster.}
\end{figure*}

We inspected the evolution of density and temperature of the simulated cluster
to determine how the temperature structure became asymmetric. Prior to the
epoch presented in Figure 7, five smaller subclusters had recently merged with
the core of the main cluster. One of these merged from the north, producing
a shock that propagated toward the south. While the shock did not leave
a strong impression on the temperature, it did disturb the ICM of the cluster.
The gas did not have time to relax before the subcluster arrived from the
south, so the gas heated by the subcluster was turbulently mixed which caused
it to spread into a large volume. The gas directly to the south was more
disturbed than the gas to the southwest, so this spreading of hot gas only
occurred on one side of the shock.

A similar scenario might explain both the asymmetry and the extent of the hot
gas near the Southern subcluster in A85. Previous mergers may have disrupted
the gas asymmetrically, causing the shock heated gas to be mixed more
efficiently on one side of the infalling subcluster. Though A85 is a cool-core
cluster, there is ample evidence from the X-ray surface brightness and
temperature that the cluster is not relaxed. Previous mergers may have
disturbed the ICM of A85 in such a way as to affect the currently observed
state of the cluster.

Figure 7 does differ in appearance from both the \textit{Chandra} and
\textit{XMM} temperature maps. The simulated cluster shows a thin layer of hot
gas on one side and extended regions of hot gas on the other, but the X-ray
temperature maps for A85 only include the extended regions. This could result
from the lower x-ray surface brightness further from the center of A85. The
X-ray count rate will tend to decrease away from the cluster center because
the density tends to decrease. Lower X-ray counts
require larger ACB or WVT regions to achieve the desired signal-to-noise for
temperature fitting. The larger the regions become, the more the temperature
will be diluted by unshocked gas and the harder it is to detect an enhanced
temperature. The ACB regions, as indicated in the top right panels of Figures
4 and 6, and the WVT regions, shown in the bottom left panels of the same
figures, are larger to the west of the Southern subcluster than they are to the
northeast. Thus, it is possible that on one side of the subcluster, smaller ACB
circles are used to extract temperatures from extended regions of hot gas,
while on the other side, larger ACB circles are used on thin layers of hot gas.
The former is not strongly affected by dilution by cooler gas since the
extraction regions are filled with heated gas, but the latter is strongly
affected.

The \textit{Enzo} simulation was used for this work to qualitatively
demonstrate how the Southern subcluster could have produced the observed
temperature structure. The simulated cluster was not intended to be
a simulation of A85; it was a part of a large sample of simulated clusters
whose ensemble properties were compared to a sample of real clusters. The
simulated cluster is only meant to be qualitatively similar to A85. The exact
details of the cluster, such as the masses, velocities, and merger trajectory,
do differ from those of A85.

\subsection{Comparison of \textit{Chandra} and \textit{XMM}}

There is a significant offset in the temperatures between \textit{Chandra} and
\textit{XMM} with \textit{Chandra} temperatures being $\sim$1 keV higher,
a large difference when the average temperature of the cluster is $\sim$6 keV.

\begin{figure*}[t!]
  \centering
  \includegraphics[height=6.35cm]{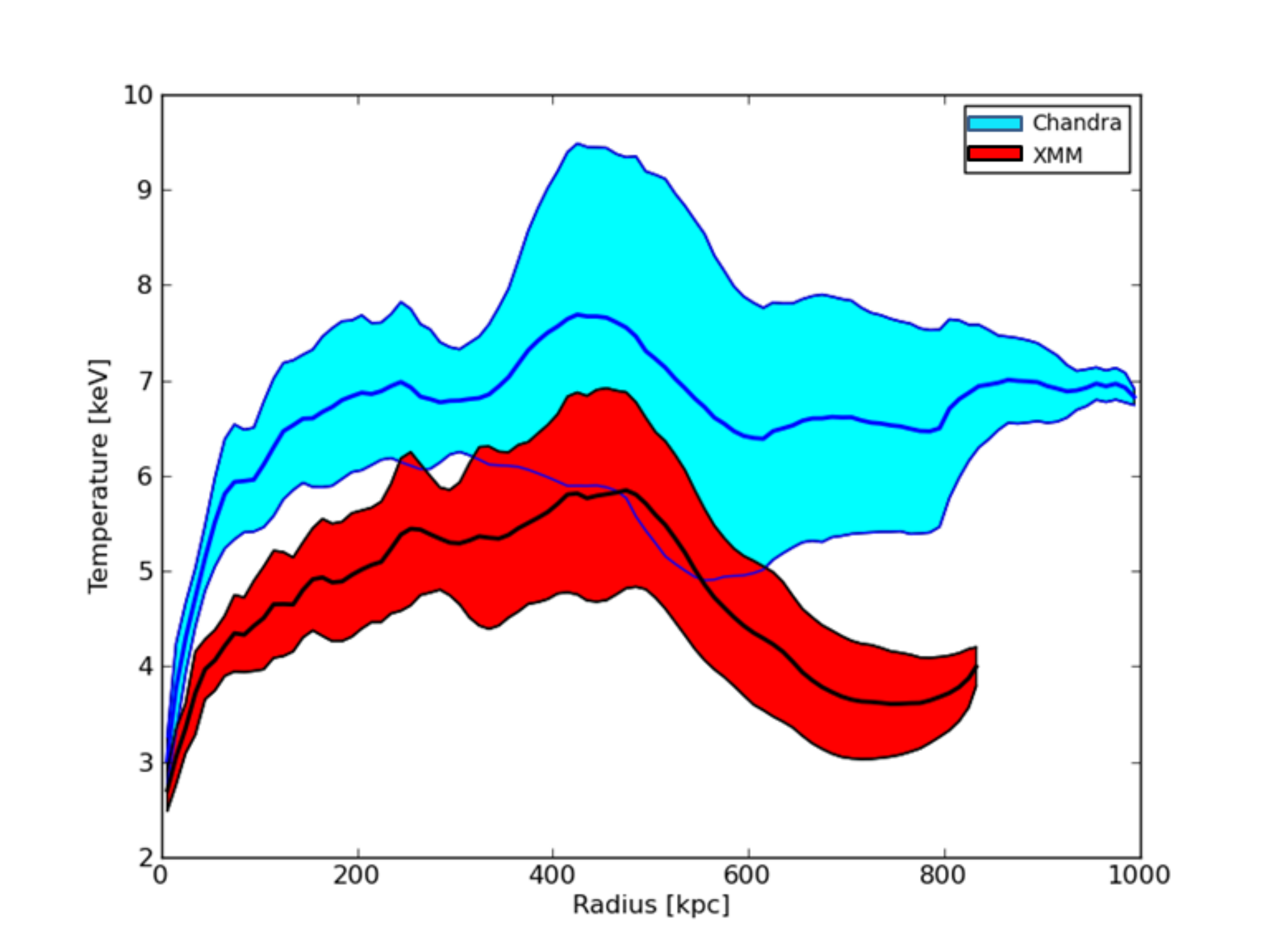}
  \includegraphics[height=6.35cm]{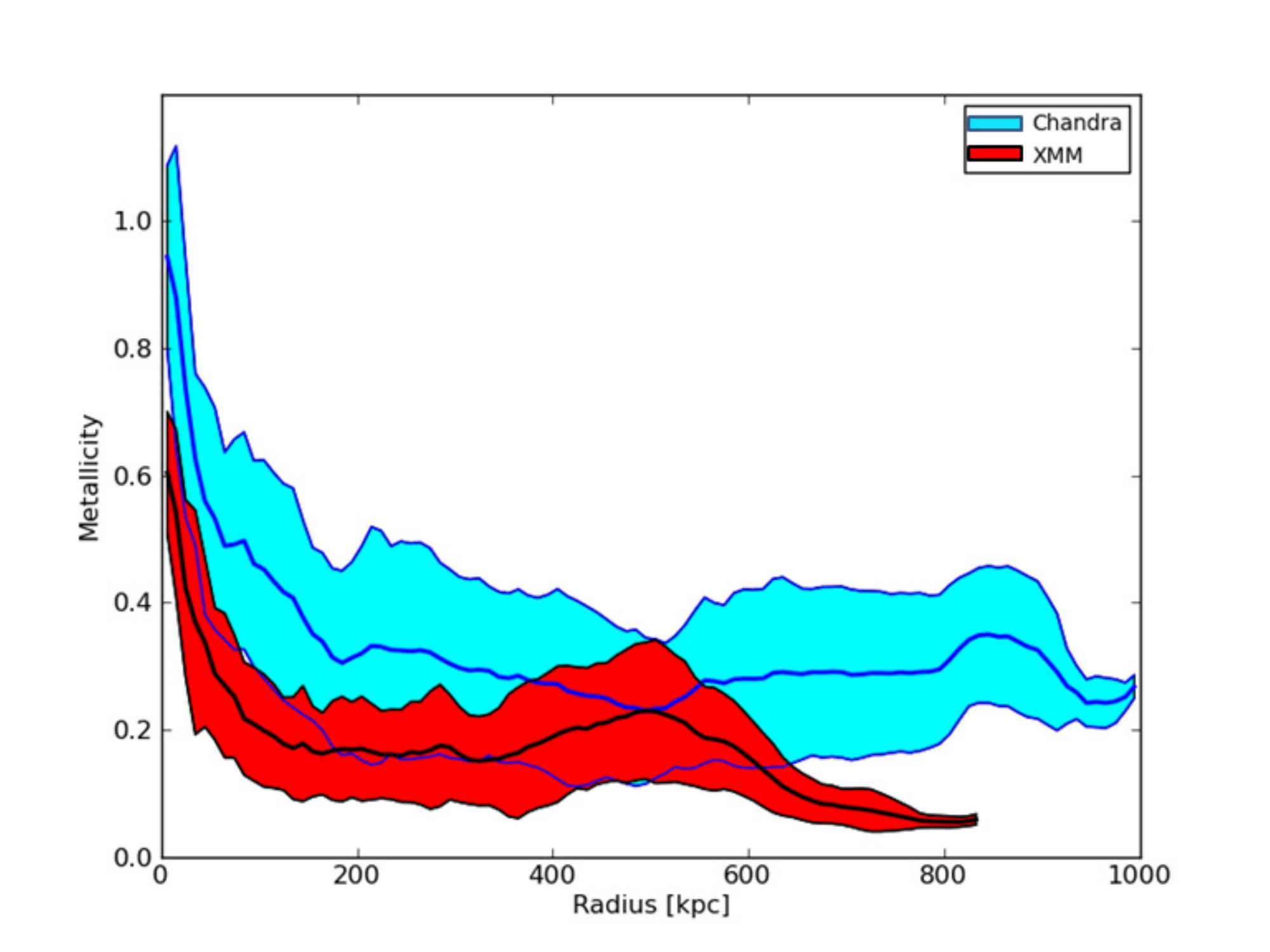}
  \caption{Left: Radial temperature profiles for \textit{Chandra} (blue) 
and \textit{XMM} (red). The \textit{Chandra} temperature exceeds the 
\textit{XMM} temperature by at least 1 keV except at the cool-core. Right: 
Radial metallicity profiles. The thick lines correspond to the mean in each 
radial bin and the shaded region equals the sample error in each bin. The 
\textit{Chandra} metallicity is approximately 50$\%$ higher at the cool-core 
and is generally higher throughout the cluster.}
\end{figure*}

Radial temperature profiles, which are given in the left panel of Figure 8,
demonstrate the offset very clearly. The radius is measured from the center of
the cool-core. The thick lines correspond to the mean temperature in a given
radial bin within the ACB map and the shaded regions correspond to the sample 
error in each bin. Except very near the cool-core, the \textit{Chandra} 
temperature exceeds the \textit{XMM} temperature by more than 1 keV.

Radial profiles were also made from the fitted metallicity and are shown in the
right panel of Figure 8. Just as with temperature, the \textit{Chandra}
metallicity is consistently higher. Both telescopes measured an enhanced
metallicity at the cool-core, but the magnitude differed between the two with
\textit{Chandra} peaking at 0.9 and \textit{XMM} peaking at 0.6.

While the individual maps displayed in Figures 4 and 6 have the same general
features such as the reduced temperature of the cool-core and the hot gas to
the northeast of the Southern subcluster, the overall level of temperature is
not in agreement and some of the finer features differ. The ACB \textit{XMM}
map includes a patch peaking at $\sim$10 keV to the west of the cluster center;
the same location in the ACB \textit{Chandra} map peaks at $\sim$8 keV. The
location of the peak temperature in the hot gas near the Southern subcluster is
separated by 2-2.5 arcminutes, or $\sim$150 kpc, between the two maps. The
temperatures near the edges of each map differ greatly, but this could be due
in part to the size of the spectral extraction regions at the edges.

The disagreement in temperature has been found in other studies seeking to
utilize the two instruments. In agreeement with our work, the magnitude of the
temperature difference was generally found to be greater for higher
temperatures and was on the order of 10-15$\%$ (Snowden et al. 2008, Nevalainen
et al. 2010, Vikhlinin et al. 2005, Schellenberger et al. 2013).

The X-ray spectra fitting procedure includes many choices that are made by the
user. We explored the effect of changing these choices to see if they altered
the temperature offset. We altered the spectral band, the size of the
extraction regions, the background scaling, the metallicity and Galactic
hydrogen column density, the emission model, and the fit statistic. While
temperature did depend on some of these factors, none of the combinations we
attempted brought the temperatures into agreement.

No satisfactory solution for the \textit{Chandra}-\textit{XMM} temperature
offset problem was discovered. We believe that the offset results from
imperfect calibration of one or both of the instruments which cannot be
corrected by the user.

\section{Shocks}

Both \textit{Chandra} and \textit{XMM} reveal regions where the temperature is higher
than the average. The highest peaks in temperature are located near the subclusters,
implying that the infall of those structures are heating the ICM. One process which may
explain the connection between the hot spots and subclusters is shock-heating. The
subclusters propagate through the gas of the main cluster and produce shocks which heat
the nearby gas.

The Mach number of a shock can be calculated using the Rankine-Hugoniot jump conditions.
The relation between the pre-shock temperature, $T_1$, post-shock
temperature, $T_2$, and Mach number, $\mathcal{M}$, is

\begin{equation}
\frac{T_2}{T_1}=\frac{(\mathcal{M}^2 +3)(5 \mathcal{M}^2 - 1)}{16 \mathcal{M}^2}.
\end{equation}

In order to calculate the Mach number, it is necessary to estimate the pre-shock temperature.
In the case of A85, it is assumed that the subclusters only disturb the main cluster ICM in
their vicinity. This is because the subclusters have a much smaller mass than A85. Far from
the subclusters, the ICM is expected to remain relaxed. If there were no subclusters present,
it is assumed that the temperature profile seen in the relaxed gas would be present throughout
the entire cluster. Thus, the pre-shock temperature is measured from regions of A85 which are
far from the hotspots. The Mach number is then estimated by comparing the temperatures of the
hot spots to the temperatures in relaxed regions which are the same distance from the cluster
center.

The hot spots in A85 are located to the South and to the West of the main cluster center. This
leaves the Northeast undisturbed. If the gas in this region truly is relaxed, then the radial
temperature profile should be statistically similar in any in sector. To test this, temperature
profiles were created along sectors shown in the left panel of Figure 9. Only the
\textit{XMM} data was used for this purpose as the field of view of the \textit{Chandra} map
was not large enough to sample the temperature in the Northeast out to the same distance as
the hot spots. The results of fitting the temperatures in these regions are shown in the right
panel of the figure. The three sectors in the Northeast, shown in the top right panel of Figure
9, do not include the hot spots, so they are expected to have the same, relaxed profile.
There are no indications that the profiles differ from one another significantly.

\begin{figure*}[t!]
  \centering
  \includegraphics[height=6.73cm]{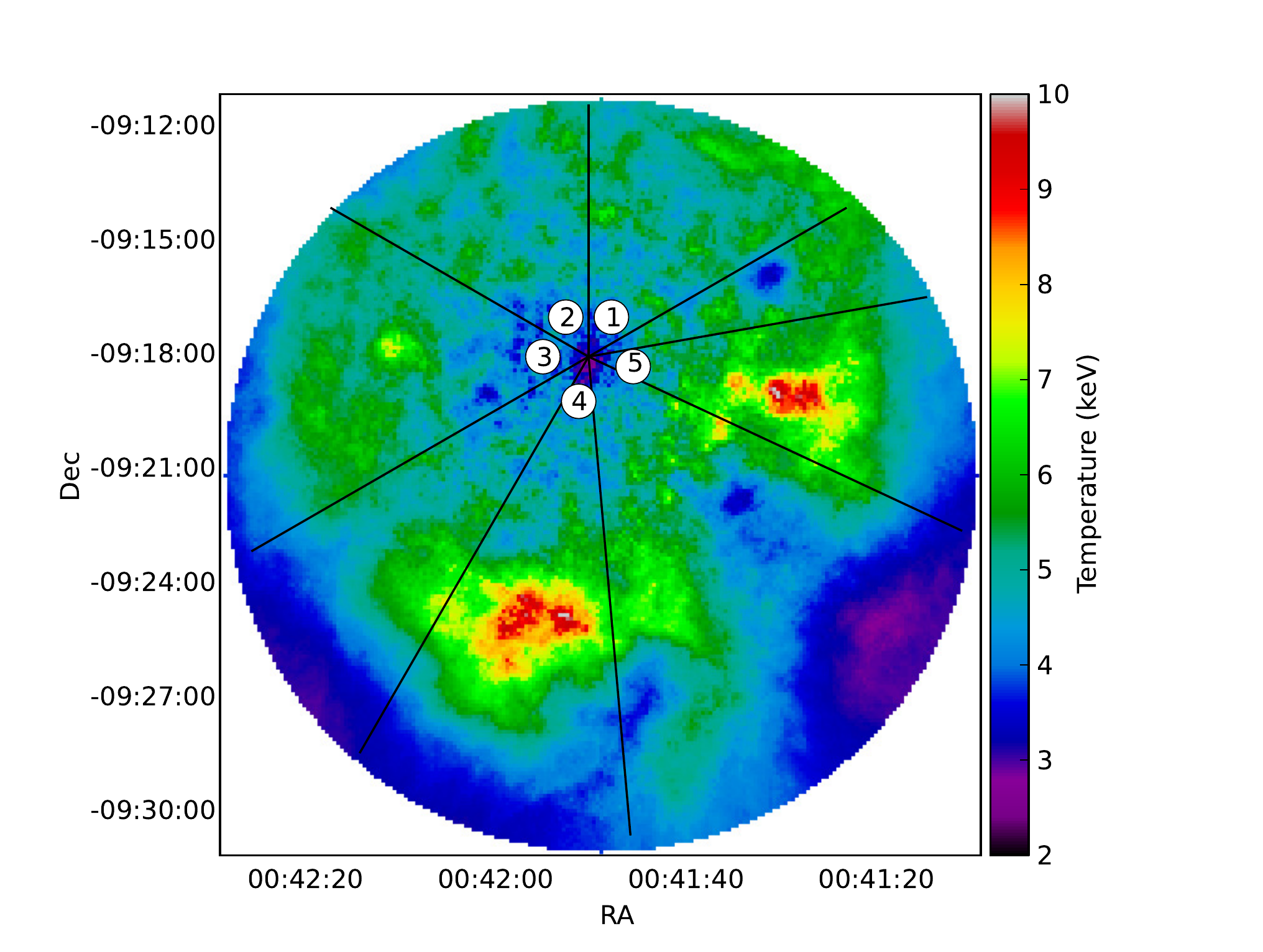}
  \includegraphics[height=6.73cm]{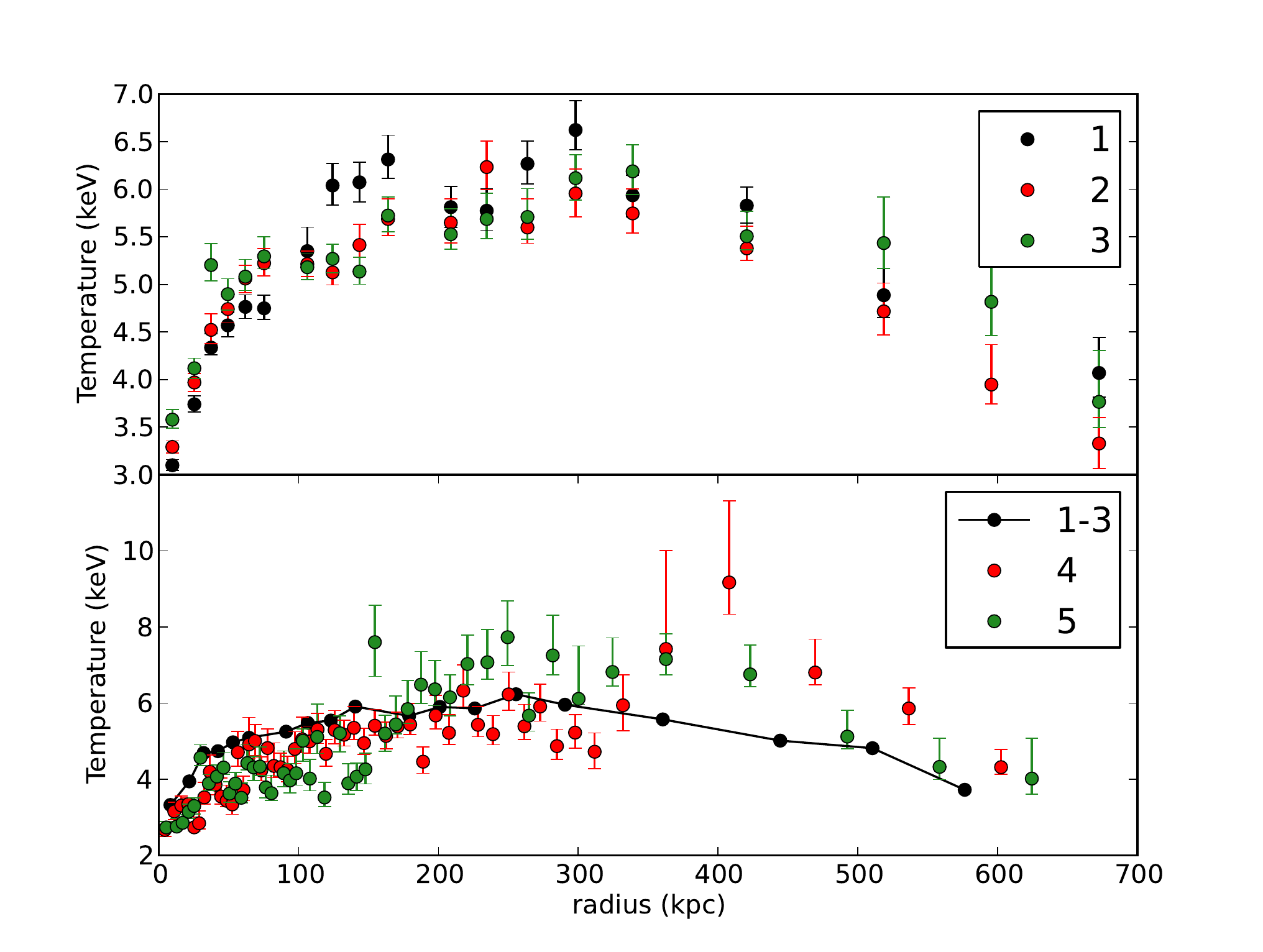}
  \caption{Left: \textit{XMM} ACB temperature map. Numbers indicate each sector from which radial
profiles were extracted. Top Right: Radial temperature profiles for sectors 1-3. These profiles reside
in the Northeast where the cluster is expected to be relaxed. Bottom Right: Radial temperature profiles
for sectors 4 and 5 as well as the average of sectors 1 through 3. The red points include the hot region
near the Southern subcluster. The hot spot is apparent at just above 400 kpc. The green points include
the hot region near the Southwestern subcluster. The heated region is seen between 150 and 300 kpc.}
\end{figure*}

The bottom right panel of Figure 9 shows the radial profiles in the direction of each
hot spot as well as the average of the three sectors believed to be relaxed. The hot spots
have significantly higher temperature than the relaxed profile at the same distance from
the cluster center. The Southern hot spot peaks at a temperature of 9.2 keV while the relaxed
temperature at the same distance is $\sim$6 keV. The peak temperature for the Western hot spot is
7.7 keV. This is lower than the peak in the ACB temperature map, which was over 10 keV, because
the spectral extraction region for the temperature profile may have been diluted by cooler gas.
Despite the lower temperature, the hot spot is still significantly higher than the relaxed
temperature profile.

Assuming the hot spots are shocks resulting from the infall of nearby subclusters, the Mach
numbers estimated from Equation 2 are $1.88 \pm 0.25$ for the Southern hot spot and
$1.91 \pm 0.24$ for the Western hot spot. These values were derived by averaging the temperatures
in radial bins in the Northeast section of the ACB map to create a relaxed temperature profile.
The peak temperatures of the hot spots were used as the post shock temperature.

Projection effects can have a strong influence on Mach number estimates from X-ray data. Both
X-ray surface brightness and temperature maps represent two-dimensional projections of the
cluster. Shocks create regions of enhanced surface brightness and temperature, but these
jumps can be diluted by unshocked gas along the line of sight. Weaker shocks covering small
volumes will be more strongly diluted than strong shocks covering large volumes as the latter
shocks will contribute a greater fraction of the flux along the line of sight. This dilution is
increased when the spectral extraction regions are large because the shock may have a filling
factor of less than unity. The point spread function of the telescope can also contribute to the
dilution.

Projection and resolution effects will tend to decrease the magnitude of temperature jumps. Thus,
the resulting Mach numbers will tend to be underestimated. This makes it very difficult to detect
low Mach number shocks in galaxy clusters, especially near the outskirts where the surface brightness
tends to be low. If the hot spots are caused by shocks, then the Mach numbers are likely systematically underestimated due to observational effects.

The technique used here can not be applied to any galaxy cluster. It requires that the gas
disturbed by the interaction between clusters only occupies a small volume, leaving a large
proportion of the ICM relaxed. For example, Abell 3667 is a major merger between two clusters
of similar size. The entire system has been disturbed and spherical symmetry is not present,
so none of the cluster is relaxed (Finoguenov et al. 2010, Datta et al. 2014). The technique
is only applicable to A85 because the mass of the main cluster is significantly higher than
the infalling sublusters.

The technique presented here assumes that the temperature profile measured far from the hot spots
represents the relaxed temperature of the cluster. If there was a large-scale temperature gradient
between the Northeast and the Southwest, then the Mach numbers would be systematically
over/underestimated depending on the direction of the gradient. The consistency between the three
radial profiles in the Northeast is evidence that there is no large-scale temperature gradient, but
it cannot be ruled out.

Previously, numerical simulations have been used to study the general properties of shocks
in galaxy clusters. Skillman et al. (2008) found that different kinds of cluster interactions
are typically associated with different magnitudes of Mach number. Accretion shocks onto
clusters are associated with Mach numbers from tens to hundreds, accretion shocks onto
filaments are associated with lower Mach numbers around 4-20, and mergers occurring within a
cluster have Mach numbers between 1 and 4. These results have been verified in other studies
(Ryu et al. 2003, Pfrommer et al. 2006, Vazza et al. 2009)Thus, the ICM of a cluster is typically
dominated by shocks with low Mach numbers. This is consistent with what was measured in A85 in
which the shock-finding algorithm did not detect any shocks exceeding a Mach number of 2.

\subsection{Radio Emission in A85}

There is extended radio emission located 320 kpc from the center of A85 and
just to the west of the Southwest subcluster. The radio emitting region, which
is shown in the bottom panel of Figure 3, appears to have a thin, filamentary
structure at 1.4 GHz. Diffuse emission with a larger extent is observed at 333
MHz (Giovannini et al. 2000). The emission has a steep spectral index. The
spectral index between 333 MHz and 1.4 GHz is between -2.5 and -3.0 (Giovannini
et al. 2000) and the spectral index measured within the 1.4 GHz band is
-3.0$\pm$0.2 (Slee et al. 2001). The polarization varies between 5$\%$ and
35$\%$ depending on the location and the integrated polarization is 16$\%$.

Due to its filamentary structure, the radio emission is often referred to as
a radio relic. Radio relics tend to be thin, elongated structures in the
outskirts of galaxy clusters. The orientation and curvature of the filamentary
structure appear consistent with the intepretation that the emission is
associated with a shock caused by the Southwest subcluster. However, the
structure could also be explained as being the remnant of a tailed radio
source.

We did not detect a temperature jump at the location of the radio emission in
either the \textit{Chandra} or \textit{XMM} data. Figure 12 shows the radio
emission overlayed onto the \textit{Chandra} and \textit{XMM} temperature maps.

\begin{figure*}[t!] 
  \centering
  \includegraphics[height=10.16cm]{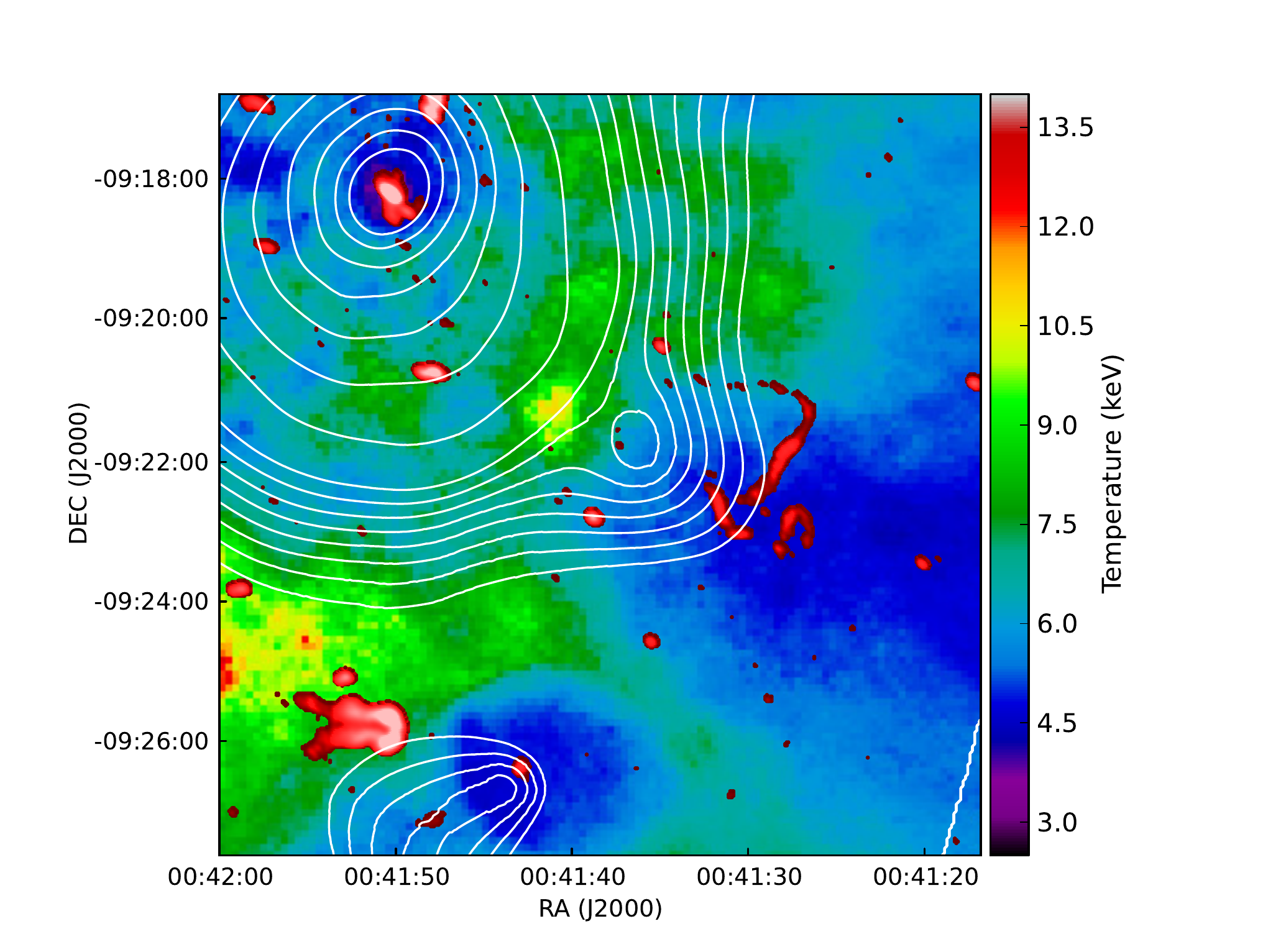}
  \includegraphics[height=10.16cm]{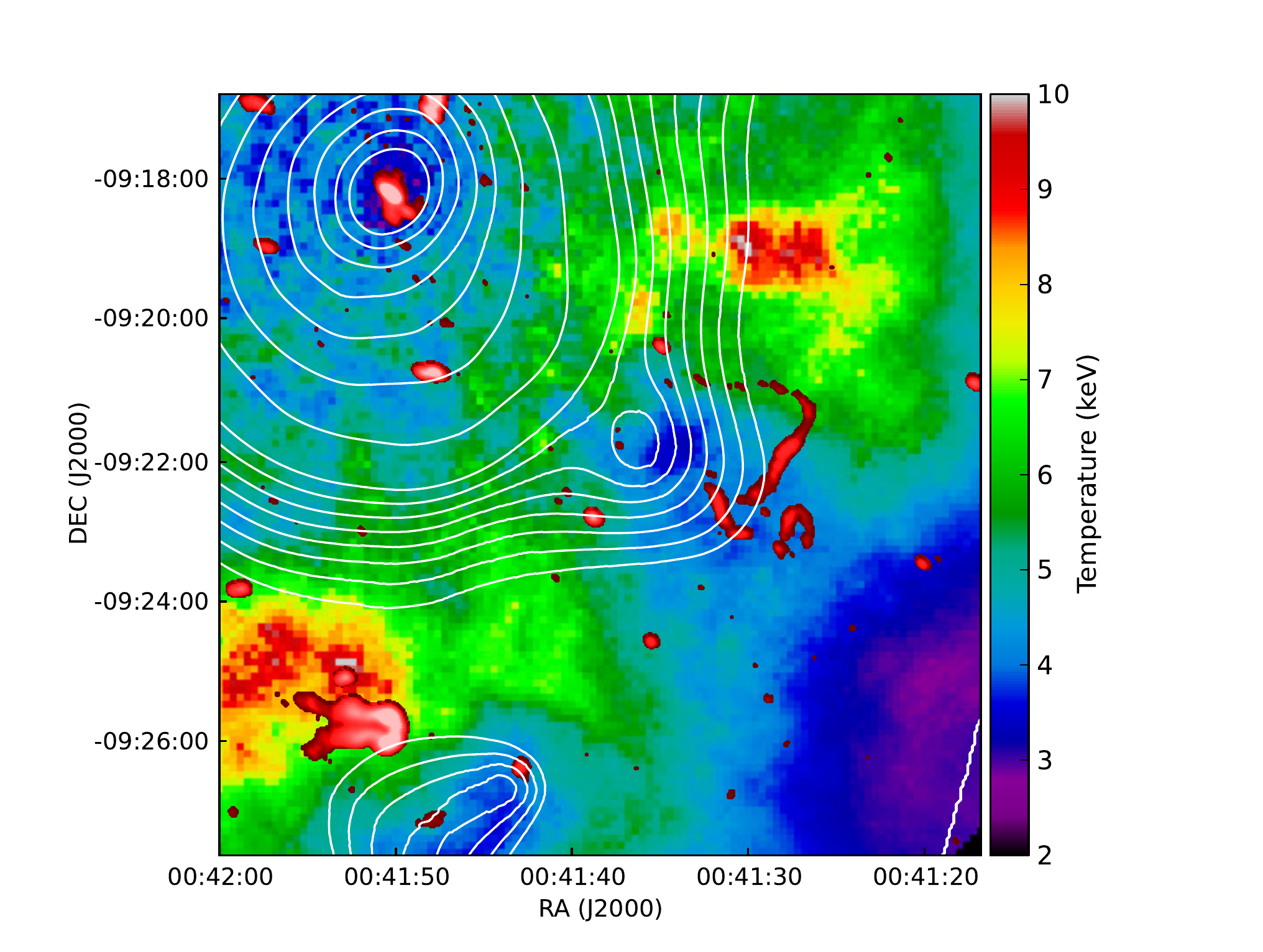}
  \caption{The 1.4 GHz radio emission measured by the VLA (red) overlayed 
onto the \textit{Chandra} temperature map (top) and the \textit{XMM} 
temperature map (bottom). The radio  emission resides in a region of cool gas.}
\end{figure*}

Despite not detecting a temperature feature at the location of the radio
emission to the Southwest, the Mach number of a potential shock can still be
estimated from the radio spectral index, $\alpha$ ($\mathcal{F}$ $\propto$
$\nu^{\alpha}$), using \begin{equation} \mathcal{M}^2
  = \frac{2\alpha-3}{2\alpha+1} \end{equation} based on diffusive shock
acceleration by a plane-parallel shock (Hoeft \& Br{\"u}ggen 2007, Ogrean et
al. 2013). Steeper radio spectra correspond to weaker shocks. Taking the
spectral index estimate of -3.0$\pm$0.2 from Slee et al. (2001), the Mach number
associated with the possible radio relic in A85 is 1.34$\pm$0.02. A shock of this
strength could be very difficult, if not impossible, to detect with the current
data due to an insufficient signal-to-noise ratio and dilution by unshocked gas.

The Mach numbers determined from radio and X-ray observations do not always
agree with one another. One example is the Toothbrush cluster in which the Mach
number of a shock was estimated to be 3.3-4.6 based on the radio spectral
index, but was no greater than 2 based on \textit{XMM} temperatures (Ogrean et
al. 2013). It is possible that dilution by unshocked gas caused the X-ray Mach
number to be underestimated. There are alternative explanations for the
difference. X-ray observations by \textit{Chandra} and \textit{XMM} may be
limited to lower Mach numbers because of the inability to measure very high
temperatures. The use of radio spectral index does not suffer from this limitation
and may be better suited at detecting high Mach number shocks.

The relative locations of the X-ray temperature features, the Southwest
subcluster, and the nearby radio emission are difficult to interpret. The radio
structure is located to the west of the subcluster, possibly indicating that
the subcluster is moving in that direction, away from the main cluster. In this
interpretation, the radio emission could result from a shock preceding the
subcluster. However, the \textit{Chandra} temperature map includes a hot spot
just to the east of the subcluster. This could also indicate the presence of
a shock, implying that the subcluster is falling in toward the main cluster.
The \textit{XMM} temperature map did not detect a hot spot at that location.

If the radio emission is not associated with a shock, it could result from the
remnants of a dead radio galaxy. Radio emission from both shocked gas and dead
radio galaxies can have a steep spectrum and high polarization fraction, which
is what is seen in the radio emission in A85. We did not find an optical
counterpart coincident with the radio emission in the SDSS image, so the source
of the radio plasma is either undetected in the optical or has moved out of the
radio emitting region. Slee et al. (2001) identified several nearby galaxies
that could have been associated with the radio emission, but they were not able
to make a decisive claim about a single galaxy based on the available data.

The origin of the radio emission to the Southwest of A85 remains unclear. If
the emission is the result of a merger shock, then a very long exposure X-ray
observation is required to detect the associated X-ray surface brightness and
temperature features.

\section{Conclusions}

We studied the temperature structure of the cool-core cluster Abell 85 using
X-ray data from \textit{Chandra} and \textit{XMM}. We used two different
methods for creating the temperature maps. The method we call Adaptive Circular
Binning produced an improved temperature map compared to Weighted Voronoi
Tesselations. Using this method on a new 100 ks \textit{XMM} observation
produced a new high resolution temperature map of A85.

The temperatures obtained from \textit{Chandra} and \textit{XMM} generally
differed from one another by 1 keV or more, especially at higher temperature.
The location of temperature structures was also not in perfect agreement
between the two telescopes, though the general structure was similar. We
adjusted our data reduction and temperature fitting procedures in search of the
origin of the temperature offset, but did not discover any user-controlled
input that brought the temperatures into agreement. We believe that the offset 
is not a result of our analysis method and is intrinsic to the instruments.

We detected asymmetric temperature substructures near the Southern and
Southwest subclusters in Abell 85 in both \textit{Chandra} and \textit{XMM}
temperature maps. These regions may be associated with merger shocks created by
the nearby subclusters. Abell 85 contains a cool-core, yet it shows evidence for
an unrelaxed ICM as the temperature structure is not symmetric or smoothly varying.

Assuming the hot regions in the temperature maps are associated with shocks, the
Mach numbers near the Southern and Southwest subclusters were 1.88 $\pm$ 0.25 and
1.91 $\pm$ 0.24, respectively, as calculated from \textit{XMM}. A similar analysis
could not be conducted using \textit{Chandra} because the observation did not have
a large enough field of view. The estimates are subject to systematic effects such
as dilution of temperature structure by unshocked gas along the line of sight. The
Mach numbers within the cluster never exceeded 2, in agreement with expecations based
on cosmological simulations which predict weak shocks dominate the ICM of clusters
(Skillman et al 2011).

We did not detect any evidence for a shock at the location of the radio
emission near the Southwest subcluster in either the X-ray surface brightness
or the temperature. However, the Mach number expected from the radio spectral
index is only 1.34. A shock this weak might not be detectable in either
surface brightness or temperature due to insufficient signal. If the relic was
caused by a shock, then much deeper observations are necessary to detect it.
If the relic is not associated with a shock, it could be the remnant of a
tailed-radio source. A more sophisticated method for modeling cluster observations
will be made available in the near future (ZuHone and Hallman, private communication).
These simulated observations will be used in future work to help determine the
sensitivity of X-ray telescopes to temperature and X-ray surface brightness features.

\acknowledgements{Acknowledgments:} The authors thank Eric Hallman, Maxim
Markevitch, Alexey Vikhlinin, Scott Randall, Steve Allen, and Norbert Werner
for taking the time to discuss data reduction and for general advice on
interpretation. We also thank the referee for their valuable input. This work
was funded by NSF grant AST-1106437 to JB. S.W.S. was partially supported by a
DOE Computational Science Graduate Fellowship under grant number DE-FG02-97ER25308.



\begin{references}

\reference{}Akamatsu, H., \& Kawahara, H.\ 2011, arXiv:1112.3030
\reference{}Bravo-Alfaro, H., Caretta, C.~A., Lobo, C., Durret, F., \& Scott, T.\ 2009, \aap, 495, 379
\reference{}Burns, J.~O., Hallman, E.~J., Gantner, B., Motl, P.~M., \& Norman, M.~L.\ 2008, \apj, 675, 1125
\reference{}Carter, J.~A.,\& Read, A.~M.\ 2007, \aap, 464, 1155
\reference{}Cash, W.\ 1979, \apj, 228, 939
\reference{}Chen, Y., Reiprich, T.~H., B{\"o}hringer, H., Ikebe, Y., \& Zhang, Y.-Y.\ 2007, \aap, 466, 805
\reference{}Datta, A., Schenck, D., Burns, J.~O. (in prep)
\reference{}Diehl, S., \& Statler, T.~S.\ 2006, \mnras, 368, 497
\reference{}Durret, F., Forman, W., Gerbal, D., Jones, C., \& Vikhlinin, A.\ 1998, \aap, 335, 41
\reference{}Durret, F., Lima Neto, G.~B., \& Forman, W.\ 2005, \aap, 432, 809
\reference{}Ferrari, C., Govoni, F., Schindler, S., Bykov, A.~M., \& Rephaeli, Y.\ 2008, \ssr, 134, 93
\reference{}Finoguenov, A., Sarazin, C.~L., Nakazawa, K., Wik, D.~R., \& Clarke, T.~E.\ 2010, \apj, 715, 1143
\reference{}Giovannini, G., \& Feretti, L.\ 2000, New A, 5, 335
\reference{}Henning, J.~W., Gantner, B., Burns, J.~O., \& Hallman, E.~J.\ 2009, \apj, 697, 1597
\reference{}Hoeft, M., \& Br{\"u}ggen, M.\ 2007, \mnras, 375, 77
\reference{}Hudson, D.~S., Mittal, R., Reiprich, T.~H., et al.\ 2010, \aap, 513, A37
\reference{}Hwang, H.~S., \& Lee, M.~G.\ 2008, \apj, 676, 218
\reference{}Kempner, J.~C., Sarazin, C.~L., \& Ricker, P.~M.\ 2002, \apj, 579, 236
\reference{}Kravtsov, A.~V., \& Borgani, S.\ 2012, \araa, 50, 353 
\reference{}Lima Neto, G.~B., \& Durret, F.\ 2003, Matter and Energy in Clusters of Galaxies, 301, 525
\reference{}Lima Neto, G.~B., Pislar, V., \& Bagchi, J.\ 2001, \aap, 368, 440
\reference{}Mahdavi, A., Hoekstra, H., Babul, A., et al.\ 2013, \apj, 767, 116
\reference{}Markevitch, M., \& Vikhlinin, A.\ 2007, \physrep, 443, 1
\reference{}Markevitch, M., Forman, W.~R., Sarazin, C.~L., \& Vikhlinin, A.\ 1998, \apj, 503, 77
\reference{}McIntosh, D.~H., Rix, H.-W., \& Caldwell, N.\ 2004, \apj, 610, 161
\reference{}Navarro, J.~F., Frenk, C.~S., \& White, S.~D.~M.\ 1996, \apj, 462, 563
\reference{}Nevalainen, J., David, L., \& Guainazzi, M.\ 2010, \aap, 523, A22
\reference{}O'Shea, B. W., Bryan, G., Bordner, J. et al. 2005a, Adaptive Mesh Refinement: Theory and Applications, ed. T. Plewa, T. Linde, \& V. G. Weirs (Berlin: Springer), 341
\reference{}Ogrean, G.~A., Br{\"u}ggen, M., van Weeren, R.~J., et al.\ 2013, \mnras, 433, 812
\reference{}Owers, M.~S., Couch, W.~J., \& Nulsen, P.~E.~J.\ 2009, \apj, 693, 901
\reference{}Pfrommer, C., Springel, V., En{\ss}lin, T.~A., \& Jubelgas, M.\ 2006, \mnras, 367, 113
\reference{}Randall, S., Nulsen, P., Forman, W.~R., et al.\ 2008, \apj, 688, 208
\reference{}Randall, S.~W., Clarke, T.~E., Nulsen, P.~E.~J., et al.\ 2010, \apj, 722, 825
\reference{}Reese, E.~D., Kawahara, H., Kitayama, T., et al.\ 2010, \apj, 721, 653 
\reference{}Ryu, D., Kang, H., Hallman, E., \& Jones, T.~W.\ 2003, \apj, 593, 599
\reference{}Sanders, J.~S.\ 2006, \mnras, 371, 829
\reference{}Santos-Lleo, M., Schartel, N., Tananaum, H., Tucker, W., \& Weisskopf, M.~C.\ 2009, \nat, 462, 997
\reference{}Schellenberger, G., Reiprich, T., Lovisari, L., Chandra-XMM Newton Cross Calibration using HIFLUGCS, IACHEC Meeting 2013, http://web.mit.edu/iachec/meetings/2013/Presentations/Schellenberger.pdf
\reference{}Skillman, S.~W., O'Shea, B.~W., Hallman, E.~J., Burns, J.~O., \& Norman, M.~L.\ 2008, \apj, 689, 1063
\reference{}Skory, S., Hallman, E., Burns, J.~O., et al.\ 2013, \apj, 763, 38
\reference{}Slee, O.~B., Roy, A.~L., Murgia, M., Andernach, H., \& Ehle, M.\ 2001, \aj, 122, 1172
\reference{}Snowden, S.~L., Mushotzky, R.~F., Kuntz, K.~D., \& Davis, D.~S.\ 2008, \aap, 478, 615
\reference{}Tanaka, N., Furuzawa, A., Miyoshi, S.~J., Tamura, T., \& Takata, T.\ 2010, \pasj, 62, 743
\reference{}Vazza, F., Brunetti, G., \& Gheller, C.\ 2009, \mnras, 395, 1333
\reference{}Vikhlinin, A., Markevitch, M., Murray, S.~S., et al.\ 2005, \apj, 628, 655

\end{references}
\end{document}